# Diffusion-controlled growth and microstructural evolution of aluminide coatings

Aloke Paul

Department of Materials Engineering, Indian Institute of Science, Bangalore – 560012, India

aloke@materials.iisc.ernet.in

**Abstract**

The diffusion-controlled growth and microstructural evolution at the interface of aluminide coatings and different substrates such as Ni-base superalloys and steel are reviewed. Quantitative diffusion analysis indicates that the diffusion rates of components in the β-NiAl phase increases with the addition of Pt. This directly reflects on the growth rate of the interdiffusion zone. The thickness and formation of precipitates between the bond coat and the superalloys increase significantly with the Pt addition. Mainly $Fe_2Al_5$ phase grows during hot dip aluminization of steel along with few other phases with very thin layer. Chemical vapor deposition process is being established for a better control of the composition of the Fe-aluminide coating on steel.

## 1. Introduction

The efficiency of jet engines with respect to the performance of turbine blades has almost reached its limit. The operating temperature, most probably, cannot be increased significantly without the discovery of a new material system. On the other hand, the effect of alloying in the existing material system is being studied extensively for the increase of service life. Turbine blades in hottest section of the jet engine are made of Ni-base superalloys. At present, these are exposed at around 0.85-0.9 of its melting point. The increase in operating temperature is the primary goal for an increased efficiency with low fuel consumption and reduction of emission of unwanted gases in the environment. The inlet gas temperature of the turbine combustor is around 1500 °C. Since the material system of a turbine blade cannot withstand such a high temperature, the gas is cooled. Additionally, these are used in a very harsh environment which needs excellent mechanical properties as well as oxidation and corrosion resistance. It is almost impossible to find a single material which will meet all the requirements. Superalloys are excellent with respect to mechanical properties. The property balance of ductility, strength, creep resistance and fracture toughness of superalloy is achieved by extensive alloying of different refractory components. However, it is still not enough with respect to the oxidation resistance. Therefore, as shown in Figure 1 [1], two different coating layers are used on the superalloy; one for the oxidation resistance (bond coat) and another for the thermal protection (top coat). At present, diffusion coating β-Ni(Pt)Al is used as a bond coat in a hotter section of a turbine blade and yittria stabilized zirconia (YSZ) is used as a top coat. The top coat is oxygen transparent. During the deposition of the top coat and service, an $Al_2O_3$ layer grows between the top and bond coat which gives protection from oxidation because of the very low diffusion rate of oxygen through it. Bond coat acts as aluminium reservoir, which supplies aluminium for the growth of the $Al_2O_3$ layer. At the same time, Al is also lost because of the growth of the interdiffusion zone in the middle of the superalloy and the bond coat. It should be noted here that superalloy is the phase mixture of γ-Ni(Al) solid solution and γ'-$Ni_3Al$ intermetallic compound. Alloying components are mainly dissolved in these two phases. An interdiffusion zone is grown between the superalloy and the bond coat because of compositional difference. The presence of refractory component rich precipitates in this zone brings a reliability issue. Therefore, an understanding of the diffusion controlled growth of this interdiffusion zone is important for further improvement of the material system in a turbine blade.

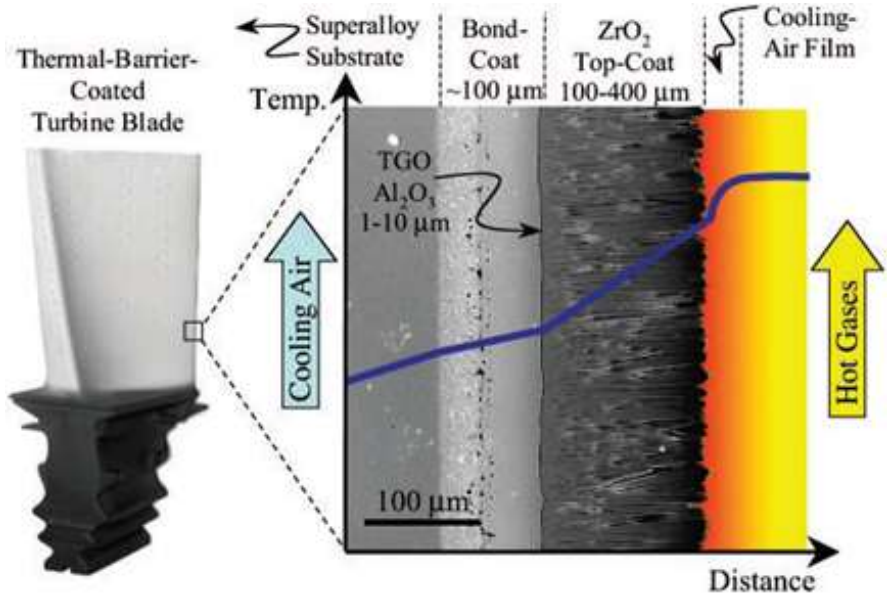

Figure 1: Cross-section of a turbine blade (adopted from Ref. [1])

Overlay MCrAlY coating is another type of bond coat used mainly in power sectors or marine applications. This is deposited directly by plasma spray process. As shown in Figure 2 [2], it has better property balance of oxidation and corrosion resistance. However, it cannot withstand high pressure and temperature like β-Ni(Pt)Al. In an aircraft engine, a superior combination of high oxidation resistance and thermal fatigue is required. In land based power generation sector and marine application, a very high corrosion resistance is required, which is achieved by the addition of Cr or a combination of Co and Cr [2]. M in MCrAlY stands for Ni and Co, which are used on Ni-base superalloys. It is basically a mixture of β and γ phases. Sometimes both Ni and Co are added depending on the requirements. In this system also, an interdiffusion zone grows because of composition difference between the bond coat and the superalloy and, therefore, similar diffusion related problems are important to discuss.

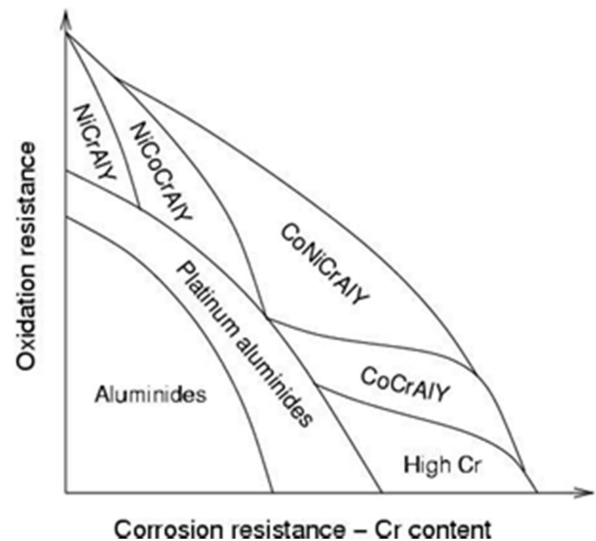

Figure 2: Comparison of oxidation and corrosion resistance of the bond coats [2].

Therefore, the aim of this review is to discuss important aspects related to the diffusion controlled growth of the phases and microstructure evolution at the coating-substrate interface. Although estimation of the diffusion parameters is discussed in detailed in other chapters of this volume, few important parameters directly relevant to the discussion in this chapter are introduced in this chapter.

Aluminide coatings are used or increasingly being proposed for the protection of steel structures in harsh environments in boilers, heat exchangers, and nuclear power plants. This is produced by different techniques such as hot dip aluminization, pack cementation, flame spraying of Al. Intermetallic compounds in the Fe-Al system grows by reaction diffusion and subsequently, an alumina layer grows on the surface at an elevated temperature, which protects the base structure in hydrocarbon and sulphur bearing environment. Number of studies available in these systems are much lower compared to the Ni-base systems. Therefore, a short description is given in the end.

## 2.1 Effect of Pt addition on growth and microstructural evolution in β-Ni(Pt)Al bond coat and Ni base superalloys

Bond coats act as a reservoir from which Al diffuses out for the growth of $Al_2O_3$ between bond coat and top coat layer during deposition of top coat and service in the application. The negligible diffusion rate of oxygen through $Al_2O_3$ makes it suitable for the protection of the substrate i.e. the superalloy from oxidation. At present, Pt-modified β−NiAl is used as a bond coat in high-pressure section, which is one of the most important sections of the jet engine bearing a combination of high temperature and pressure in a very harsh and corrosive environment. [3]. The addition of Pt helps in decreasing the sulfur segregation at one of the weakest interfaces *i.e.* the bond coat/$Al_2O_3$ interface and therefore increases the reliability by improving the adherence of the oxide layer with the bond coat. It is believed that it enhances the lifespan of the material system by growing the $Al_2O_3$ layer at a faster rate after the spallation of the oxide layer because of thermal mismatch. At the same time, the loss of Al at the superalloy and bond coat interface is a serious issue. During service, an interdiffusion zone grows at the interface of bond coat and the superalloy because of composition difference. There is an Al loss from the bond coat and Ni loss from the superalloy substrate. The refractory components are used at a very high concentration to achieve a superior mechanical property. Topologically close-packed (TCP) precipitates grow because of loss of Ni, which is undesirable. There are many articles available studying the effect of Pt on different aspects such as microstructural evolution of aluminide coatings during pack cementation [4-7], oxidation rate of the coating [8-11], quality and nature of the oxide layer [12-15], mechanism of degradation of the coating [16-18] and reasons behind rumpling of the oxide layer [19-21].

In an industry, β-(Ni,Pt)Al is grown first by electroplating of Pt on the Ni-base superalloy and then Al is deposited by pack cementation at an elevated temperature. This is explained with the help of Ni-Al phase diagram [26]. As already mentioned previously, Ni-base superalloy is a phase mixture of γ and γ' phase, as shown in Figure 3. Other alloying components are dissolved in these two phases. After electroplating a thin layer of Pt, when Al is deposited on the superalloy at a desired high temperature, Al-rich intermetallics and β-Ni(Pt)Al grow. Following, a single β-Ni(Pt)Al phase layer is developed with the desired composition distribution. During this stage itself, an interdiffusion zone between the bond coat and the superalloy starts growing, as shown in Figure 4 [27]. Therefore, there is a loss of Al on both the sides (for the growth of IDZ and $Al_2O_3$) and loss of Ni from the superalloy.

In most of the studies, the growth of the interdiffusion zone during pack cementation and oxidation studies are reported. Because of a wide composition range of the bond coat and the presence of many components, it is difficult to develop an understanding of the diffusion controlled growth mechanism of IDZ based on the quantitative analysis. Quantitative diffusion analyses are done mostly in the binary Ni-Al system [28-33] and very recently in Ni-Pt-Al system [34] following a pseudo-binary approach [35-37]. Therefore, as shown in Figure 5, the aim of this section is to make a link between quantitative diffusion studies and the growth of IDZ in real applications. Experiments on the growth of IDZ are conducted following the diffusion couple technique that mimics the real structure. Two alloys of bond coats $(Ni_{50-x}Pt_x)Al_{50}$ and $(Ni_{60-x}Pt_x)Al_{40}$ with fixed compositions are arc melted with Pt content of x = 0, 5, 10 and 15 at.%. Following, these are diffusion coupled with two different single crystal superalloys, René N5 and CMSX-4. However, to correlate the estimated diffusion coefficients with the growth rate of IDZ, the diffusion data in binary Ni-Al and ternary Ni-Pt-Al are discussed in brief.

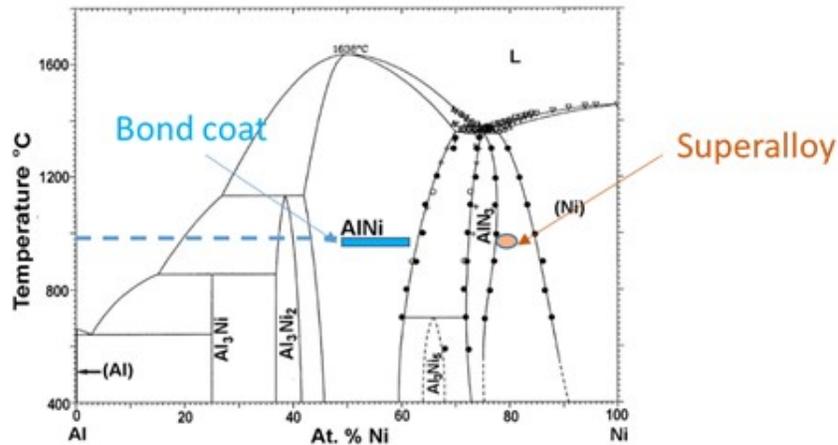

Figure 3 The growth of bond coat is explained with the help of Ni-Al phase diagram [38]

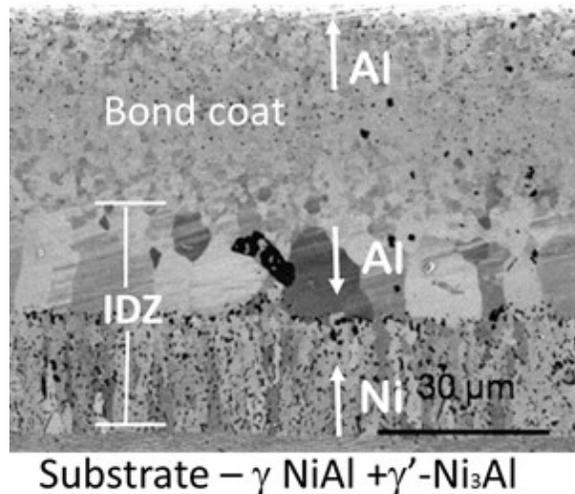

Figure 4 The growth of an interdiffusion zone (IDZ) between the superalloy substrate and the bond coat [27]

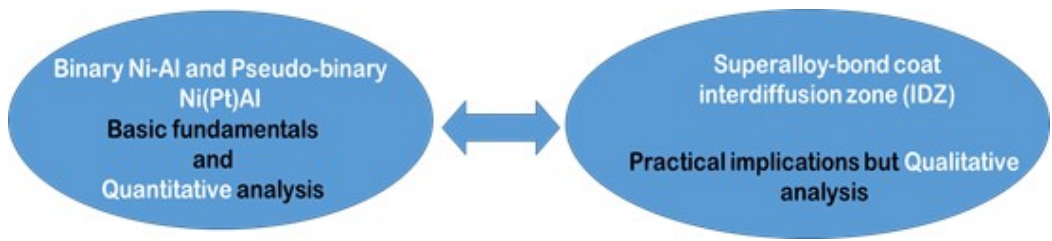

Figure 5 A schematic representation of the missing link between fundamental studies and diffusion controlled growth in real applications.

### 2.1.1 Diffusion rates of Ni and Al in β-NiAl

The estimation of diffusion coefficients in the Ni-Al system is extensively discussed in another chapter in this issue. Only very important data directly relevant to the discussion of this chapter is discussed here along with the data on the effect of Pt estimated [34] following the pseudo-binary approach [35-37]. Several incremental diffusion couples were prepared to estimate the interdiffusion and intrinsic diffusion coefficients in the binary β-NiAl phase [32, 33, 38]. One of such diffusion couple with the presence of a special phenomenon i.e. the bifurcation of the Kirkendall marker plane is shown in Figure 6. When the diffusion couple was prepared with the alloys $Ni_{72}Al_{28}$ and $Ni_{42}Al_{58}$ to at 1000 °C for 24 hrs, the Kirkendall marker plane was split into two witnessing the presence of a bifurcation of the Kirkendall marker plane which is very rare phenomenon [39]. One plane is found in the Ni-rich and another plane is found in the Al-rich side of the β-NiAl phase.

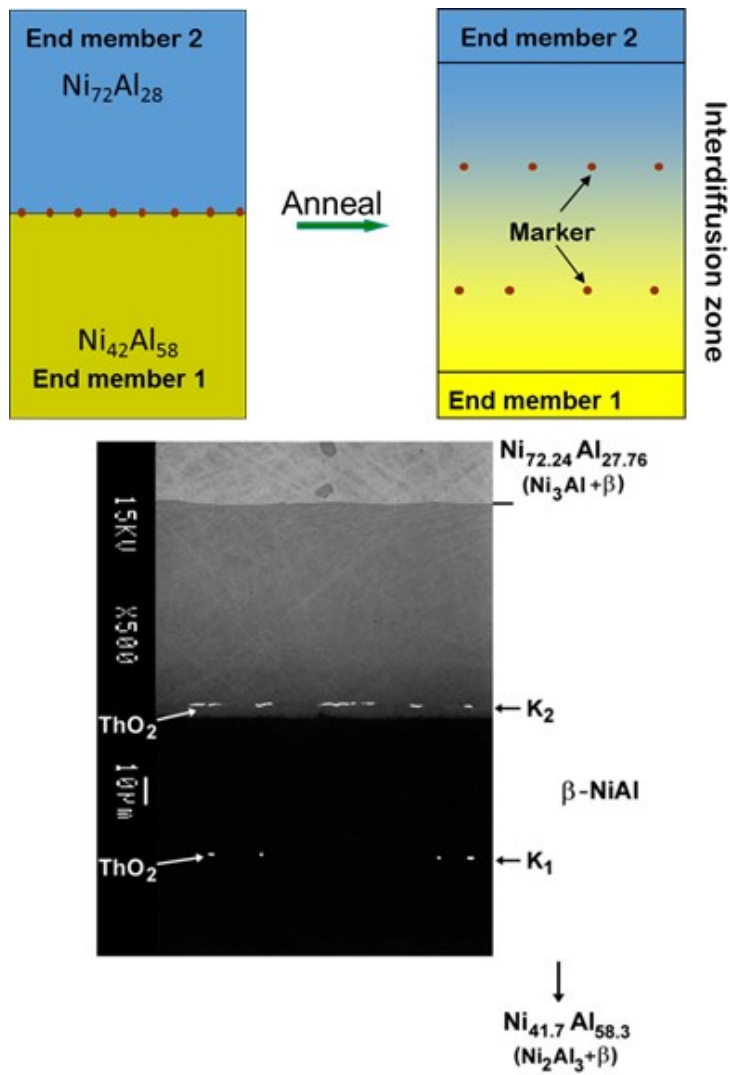

Figure 6: Bifurcation of the Kirkendall marker plane in a $Ni_{72}Al_{28}/Ni_{42}Al_{58}$ diffusion couple annealed at 1000 °C for 24 hrs. The Kirkendall marker planes $K_1$ and $K_2$ are identified with the help of $ThO_2$ powders [32].

In the process, a special microstructural feature was noticed, which led to the development of a physico-chemical approach [40-45]. It is found that the microstructural evolution is strongly dependent on the location of the marker plane. A duplex morphology is found in the case, the marker plane is found inside the same phase. If it is present at the interphase, long grains are found to cover the whole phase layer. This is found in most of the cases since the nucleation barrier for new grains to form must be high. In few other cases, the interdiffusion zone is covered by many small grains [46-47]; however, one can still locate the marker plane from the microstructural features. The same is found in the case of Ni-Al system, as shown in Figure 7. This is the same diffusion couple that is shown in Figure 6, in which only a partial area around the Kirkendall marker planes is shown. Two different microstructures around $K_1$ can be understood from the Figure 7 showing the whole interdiffusion zone. A focused area around this marker plane is shown in which different grain morphologies around $K_2$ is clear. It indicates that the grains grow differently from two different interfaces of a phase and meet at the Kirkendall marker plane [40-45]. This discussion will be brought back again during the discussion of microstructural evolution between the bond coat and the superalloy.

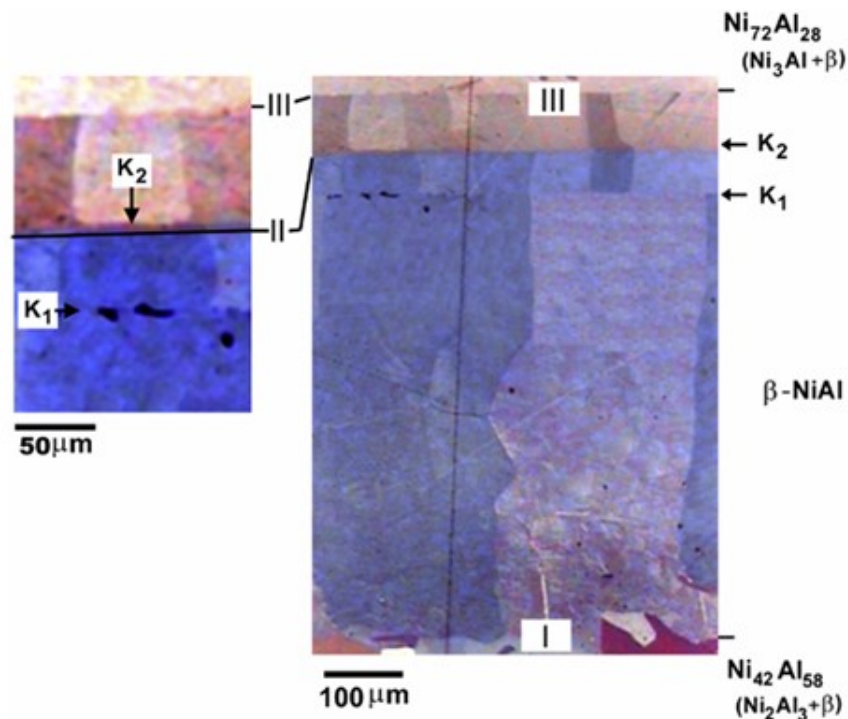

Figure 7: Microstructural feature of the interdiffusion zone of the $Ni_{72}Al_{28}/Ni_{42}Al_{58}$ diffusion couple annealed at 1000 °C for 24 hrs [43]

The measurement of the interdiffusion coefficients in this phase is explained in detail in another chapter. One can also go through the original articles. In this chapter, only the estimated results are shown in Figure 8 [32]. The data points are measured from many diffusion couples and they fall in the range of experimental error. The estimated ratio of the intrinsic diffusion coefficients is shown in Figure 9. It can be seen that the interdiffusion coefficient is minimum near the stoichiometric composition of Ni:Al = 50:50. It increases drastically with the deviation of composition from the stoichiometry. This is more prominent in the Al-rich side compared to the Ni-rich side. Secondly, as can be seen in Figure 9, Ni has higher diffusion rate compared to Al on the Ni-rich side, whereas, Al has higher diffusion rate compared to Ni in the Al-rich side. To understand these facts, we need to consider thermodynamic driving forces and defect concentrations.

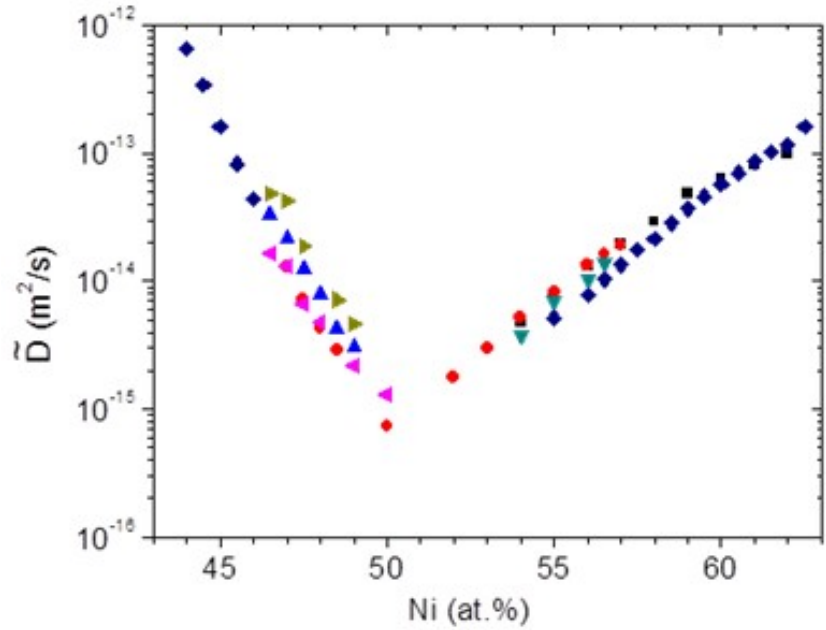

Figure 8 The variation of the interdiffusion coefficients in the β-NiAl phase [32].

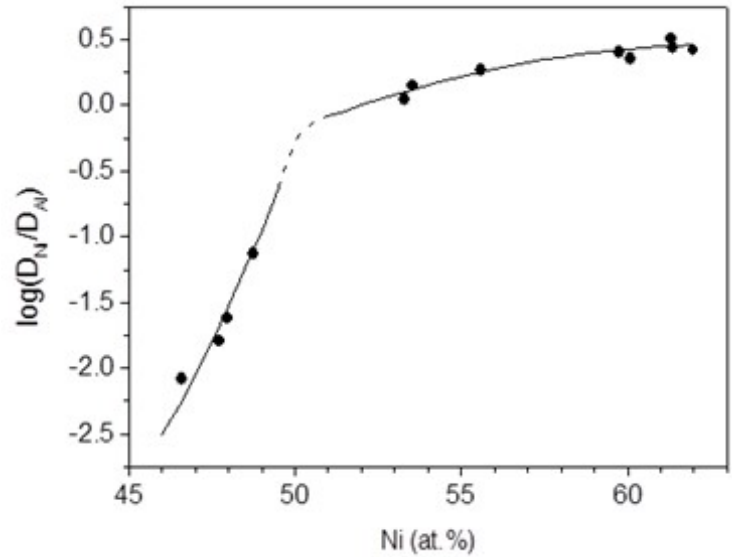

Figure 9 The ratio of the intrinsic diffusion coefficients [32].

The interdiffusion coefficient $\tilde{D}$ is related to the intrinsic ($D_i$) and tracer ($D_i^*$) diffusion coefficients by [48]

$$\tilde{D} = C_{Al}V_{Al}D_{Ni} + C_{Ni}V_{Ni}D_{Al} = (N_{Al}D_{Ni}^* + N_{Ni}D_{Al}^*)\Phi W, \qquad (1)$$

where $C_i = \frac{N_i}{V_m}$ is the concentration of component $i$, $N_i$ is the composition in atomic or mol fraction, $V_m$ is the molar volume, $V_i$ is the partial molar volume, $\Phi = \frac{dlna_{Al}}{dlnN_{Al}} = \frac{dlna_{Ni}}{dlnN_{Ni}}$ is the thermodynamic driving force and W is the vacancy wind effect. This is the same for both the components in a binary system [48]. As it is found in many systems [49], the trend of variation of interdiffusion coefficients is similar to the variation of the thermodynamic driving force on components. The variation of this parameter in the β-NiAl phase is shown in Figure 10 [32]. On the other hand, it can be seen that the trends of variation of the interdiffusion coefficients and the thermodynamic driving forces are just opposite in this system. This indicates that the concentration of defects must have more influential role over the thermodynamic driving forces to influence on the diffusion rates of the components.

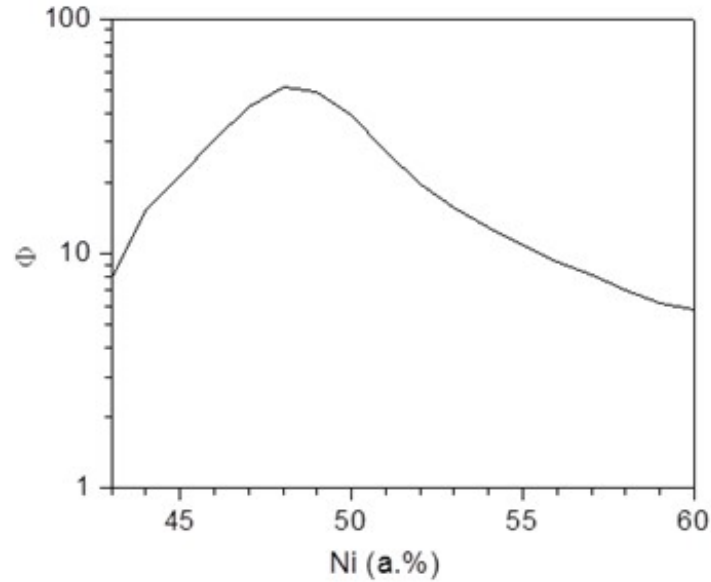

Figure 10 The variation of the thermodynamic driving force with the composition in the β-NiAl phase [49].

This β-NiAl phase has CsCl structure. As shown in Figure 11, it can be seen as two interpenetrating simple cubes of Ni and Al. Therefore, both are surrounded by unlike components. Now it is known that two types of defects are present in this phase [50]. In the Al-rich side of the phase, triple defects are present in which it has two vacancies on Ni sublattice and one Ni antisite (not necessarily associated together [50]). In the Ni-rich side, Ni antisites are present. The concentrations of these structural defects increase with the increase in deviation of composition from the stoichiometry.

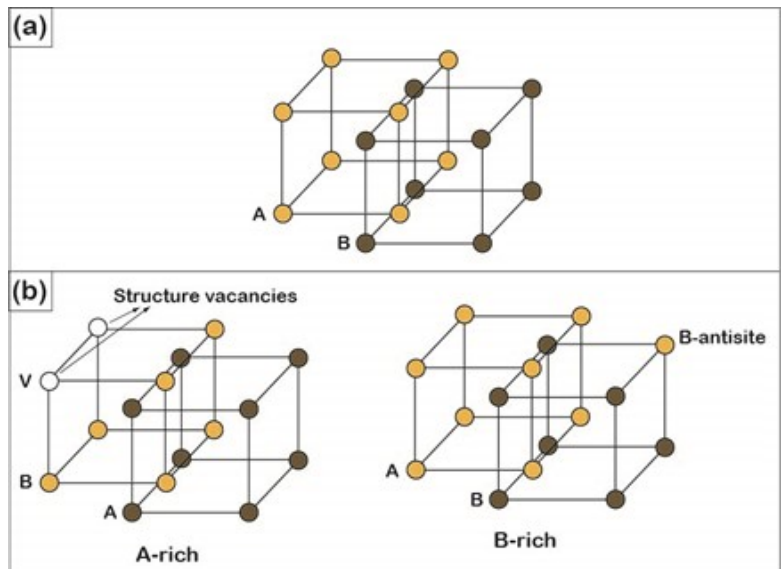

Figure 11 Types of defects in β-NiAl (a) defect free structure and (b) the presence of structural vacancies and antisites.

The calculated concentrations of defects by S. Divinski in Ref. [39] are shown in Figure 12. A very high concentration of vacancies in the Al-rich side justifies the sharp increase in interdiffusion coefficient in the Al-rich side. On the other hand, the presence of Ni antisites on the Ni-rich side explains the increase of the diffusion coefficient with a deviation of the composition from the stoichiometry. Further, the triple defect mechanism [39] in the Al-rich side explains the higher diffusion rate of Al compared to Ni. On the other hand, antisite bridge mechanism [53] is operative on the Ni-rich side and the presence of Ni-antisites explains the higher diffusion rate of Ni compared to Al on this side of the stoichiometry [53].

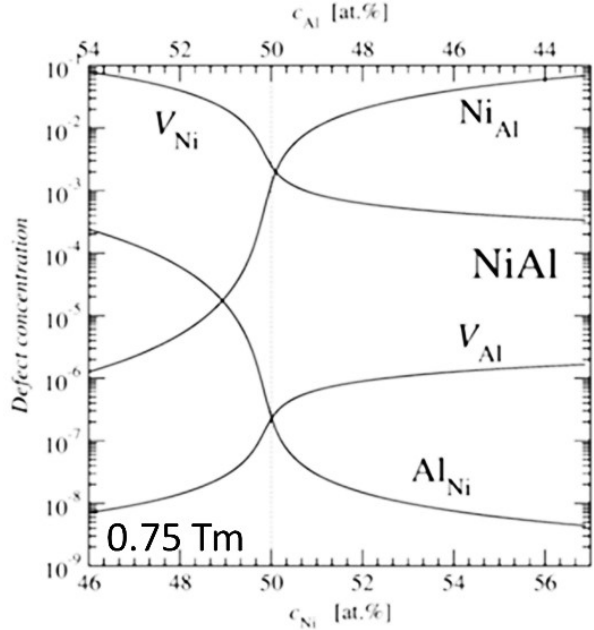

Figure 12 The variation of the defect concentrations on different sublattices in the β-NiAl phase [39].

### 2.1.2 Effect of Pt on diffusion rates of components of Ni and Al in β-Ni(Pt)Al

Based on the theoretical analysis by Marino and Carter [52] indicating the decrease in defect formation and migration energy with the Pt addition, one can expect that the diffusion rate of Ni and Al should increase. Minamino et al. [53] estimated the diffusion rate of Pt in different β-NiAl alloys. However, it should be noted here that the diffusion rates of Ni and Al because of the presence of Pt is more important to study for the sake of understanding the growth of the interdiffusion zone between the bond coat and the superalloy. There was no interdiffusion and tracer diffusion studies available examining this fact. Very recently, Kiruthika and Paul [34] conducted pseudo binary diffusion couple studies [35-37] in the β-Ni(Pt)Al phase, which is explained in this section in detail.

The advantage of following the pseudo-binary diffusion couple compared to the conventional method needs to be explained first. In a system with $n$ number of component, $\tilde{J}_i$ i.e. the interdiffusion flux of component $i$ is related to the diffusion coefficients as [39]

$$\tilde{J}_i = -\sum_{j=1}^{n-1} \tilde{D}_{ij}^n \frac{\partial C_j}{\partial x}, \qquad (2)$$

where the concentration gradient is expressed as $\dfrac{\partial C}{\partial x}$. Here $n^{th}$ component is considered as the dependent variable and therefore we have $(n\text{-}1)$ interdiffusion coefficients. The interdiffusion fluxes are related by

$$\sum_{i=1}^{n} V_i \widetilde{J}_i = 0 \qquad (3)$$

The interdiffusion fluxes of a component can be estimated by den Broeder relation by [39]

$$\widetilde{J}_B = \frac{C_B^+ - C_B^-}{2t}\left[\left(1 - Y_C^*\right)\int_{x^{-\infty}}^{x^*} Y_C\, dx + Y_C^* \int_{x^*}^{x^{+\infty}}\left(1 - Y_C\right) dx\right] \qquad (4)$$

$Y_C = \dfrac{C_B - C_B^-}{C_B^+ - C_B^-}$ is the concentration $\left(C_B = \dfrac{N_B}{V_m}\right)$ normalized variable. "*" identifies the parameters at the point of interest. Subsequently, the interdiffusion coefficients are estimated applying the Fick's law as expressed in Equation 1. The interdiffusion fluxes of components are estimated utilizing the Equation 3 in a binary or multicomponent system. The number of interdiffusion coefficients to be estimated depends on the number of components in a system.

In a binary system ($n = 2$), the interdiffusion fluxes of the components are related by interdiffusion coefficient (following Equation 1a)

$$\widetilde{J}_1 = -\widetilde{D}\frac{dC_1}{dx}\,;\;\; \widetilde{J}_2 = -\widetilde{D}\frac{dC_2}{dx} \qquad (5a)$$

$$V_1 \widetilde{J}_1 + V_2 \widetilde{J}_2 = 0 \qquad (5b)$$

Therefore, it has only one interdiffusion coefficient in a binary system. In a ternary system ($n = 3$), these can be expressed as

$$\widetilde{J}_1 = -\widetilde{D}_{11}^3 \frac{\partial C_1}{\partial x} - \widetilde{D}_{12}^3 \frac{\partial C_2}{\partial x} \qquad (6a)$$

$$\widetilde{J}_2 = -\widetilde{D}_{21}^3 \frac{\partial C_1}{\partial x} - \widetilde{D}_{22}^3 \frac{\partial C_2}{\partial x} \qquad (6b)$$

$$V_1 \widetilde{J}_1 + V_2 \widetilde{J}_2 + V_3 \widetilde{J}_3 = 0\,. \qquad (6c)$$

It can be seen that only two interdiffusion fluxes need to be determined and the third one can be estimated from these two values. $\widetilde{D}_{11}^3$ and $\widetilde{D}_{22}^3$ are the main interdiffusion coefficients and $\widetilde{D}_{12}^3$ $\widetilde{D}_{21}^3$ are the cross interdiffusion coefficients [39]. Component 3 is the dependent variable [39]. In ternary or multicomponent systems, in general, the lattice parameters with the variation of the composition are not known and therefore a constant molar volume is considered for the analysis. In such a situation, the equations written above can be expressed as

$$\widetilde{J}_1 = -\widetilde{D}_{11}^3 \frac{1}{V_m}\frac{\partial N_1}{\partial x} - \widetilde{D}_{12}^3 \frac{1}{V_m}\frac{\partial N_2}{\partial x} \qquad (7a)$$

$$\widetilde{J}_2 = -\widetilde{D}_{21}^3 \frac{1}{V_m} \frac{\partial N_1}{\partial x} - \widetilde{D}_{22}^3 \frac{1}{V_m} \frac{\partial N_2}{\partial x} \tag{7b}$$

$$\widetilde{J}_1 + \widetilde{J}_2 + \widetilde{J}_3 = 0 . \tag{7c}$$

It can be understood from the equations that two interdiffusion fluxes can be estimated from a diffusion couple. On the other hand, there are four interdiffusion coefficients to estimate in a ternary system. Therefore, we need two diffusion couples for the estimation of these parameters. Since the diffusion parameters are material constants (if different diffusion couple does not follow different diffusion mechanisms), these parameters can be estimated at the composition of the intersection of two different diffusion couples. However, it brings a drawback for a systematic study, which is explained in Figure 13. Suppose a set of diffusion couples are prepared to estimate the diffusion parameters at one particular (Ni+Pt): Al composition. With the aim of studying the effect of Pt addition on diffusion rates of components suppose another set of diffusion couples are prepared by changing the Pt content in the end members. However, there is a possibility that the composition of the intersection will have a different ratio of (Ni+Pt): Al (as shown by solid and dotted lines) because of serpentine nature of the diffusion profiles on Gibb's triangle [39].  Therefore, data generated in ternary systems are mostly random unless these are extrapolated to develop a trend in change in diffusivities with composition [54]. In a system, with a higher number of components, one cannot estimate these parameters because of mathematical complications and average interdiffusion coefficients are estimated. Intrinsic diffusion coefficients are also not possible to estimate and difficult to correlate the estimated data for the understanding of the atomic mechanism of diffusion [39].

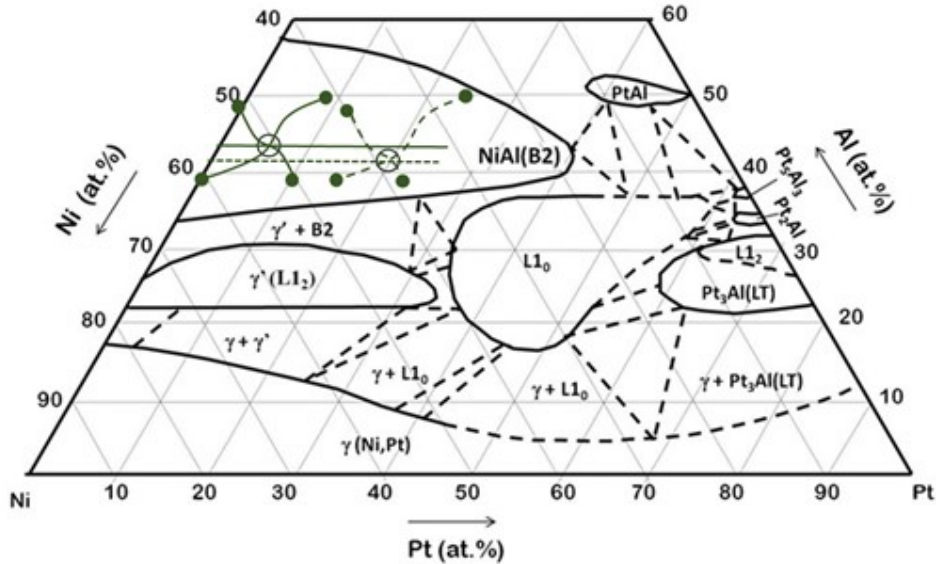

Figure 13: Explanation of the difficulties of estimation of diffusion parameters with composition systematically in a ternary Ni-Pt-Al system (the phase diagram is adopted from [26]).

To circumvent this problem, the present author proposed a pseudo-binary approach for the systematic estimation of the diffusion coefficients. The main advantages are:

(i)     One can estimate the variation of the interdiffusion coefficients with composition from a single diffusion couple [34, 36].

(ii)    It is possible to estimate the intrinsic diffusion coefficients at the marker plane (similar to the binary system) [36]

(iii)    It is possible to study the effect of alloying on diffusion rates of components systematically [34].

(iv)    One can even relate to the atomic mechanism of diffusion in correlation with thermodynamic parameters and predict the effect of alloying on defects assisting the diffusion process [37].

(v)    When combined with the physico-chemical approach [39, 42], one can relate the change in diffusion rates of components with microstructural evolution in a multicomponent multiphase interdiffusion zone.

However, one of the drawbacks of this method is that one cannot estimate the cross diffusion coefficients, which sometimes have bigger influence compared to the main diffusion coefficients [54].

Two sets of alloys were produced for making the pseudo−binary diffusion couples in β-Ni(Pt)-Al phase, as shown in Figure 14: $Ni_{60-x}Pt_xAl_{40}$ and $Ni_{50-x}Pt_xAl_{50}$, x = 5, 10 and 15 at.%. These diffusion couples were annealed at 1100 °C for 25 h. Further details can be found in Ref [49].

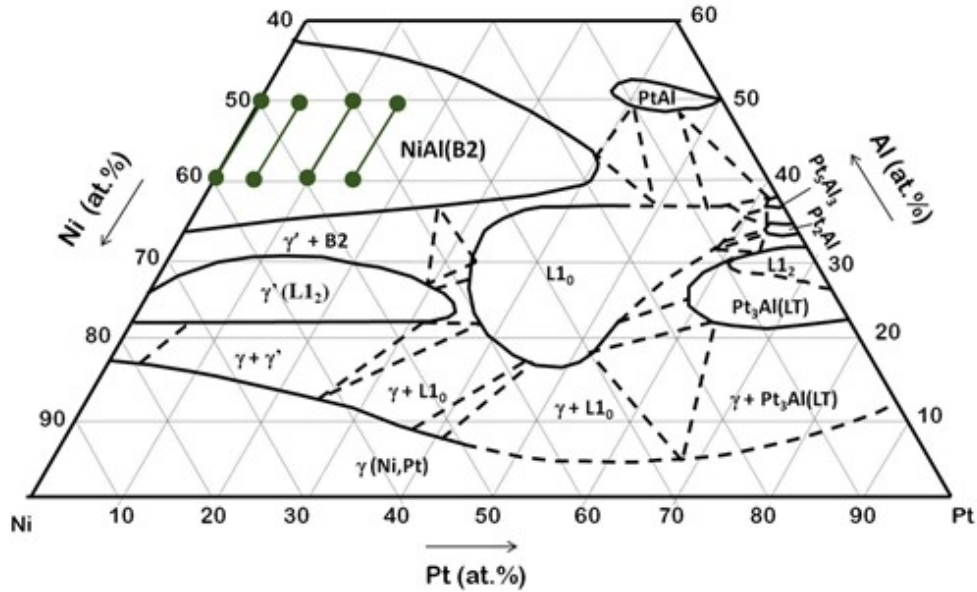

Figure 14: Compositions of the end members for pseudo-binary diffusion couples in Ni-Pt-Al system (the phase diagram is adopted from [26]).

An example of the composition profile of the pseudo-binary diffusion couple $Ni_{60}Al_{40}/Ni_{50}Al_{50}$ in this phase is shown in Figure 15 [34]. The composition of Pt remains more or the less, which makes it suitable for the estimation of the data following this approach. The variation of activities of the components in the diffusion couple, as shown in Figure 16 justifies the reason for developing such a profile, since the variation of the activity of Pt is negligible throughout thus couple [34]. It should be noted here that this method should not be followed in the presence of a prominent uphill diffusion [37], which was done by Tsai et al. [55] in high entropy alloys. In such a situation, one can estimate the average interdiffusion coefficients for different components instead of estimation of composition dependent interdiffusion coefficients, which have no physical significance.

Following a similar analysis as explained in Ref [37] Equations 6 and 7 reduces to

$$\tilde{J}_{Ni} = -\tilde{D}\frac{dc_{Ni}}{dx} \qquad (8a)$$

$$\tilde{J}_{Al} = -\tilde{D}\frac{dC_{Al}}{dx} \qquad (8b)$$

$$\tilde{J}_{Ni} + \tilde{J}_{Al} = 0 \qquad (8c)$$

$$N_{Ni} + N_{Al} + N_{Pt} = 1 \qquad (8d)$$

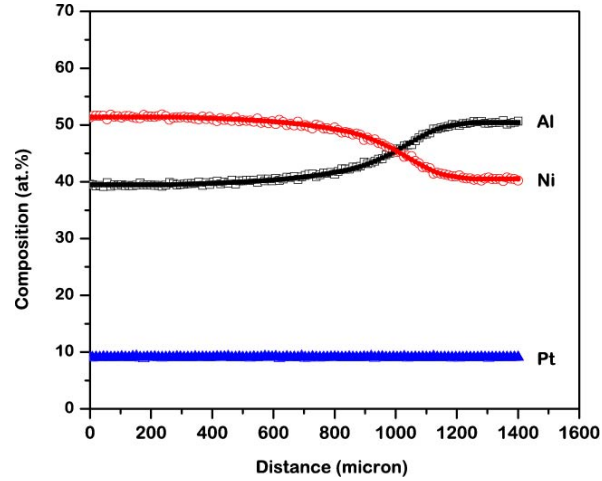

Figure 15 A measured composition profile of pseudo-binary $Ni_{60}Al_{40}/Ni_{50}Al_{50}$ diffusion couple annealed at 1100 $^{o}$C for 25 h [34].

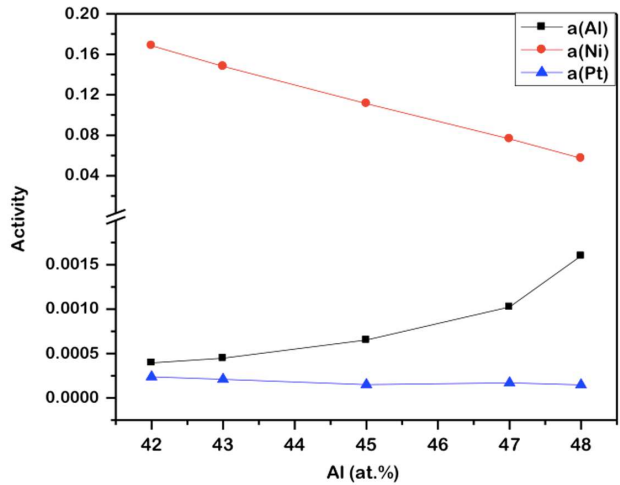

Figure 16 Variation of activities of components in $Ni_{60}Al_{40}/Ni_{50}Al_{50}$ diffusion couple [34].

Therefore, we will have a same value of the interdiffusion coefficient when we estimate following the composition profile of Ni or Al. For the sake of estimation and since Pt replaces Ni at the same sublattice of Ni, the compositions should be normalized as [34]

$$M_{Ni} = N_{Ni} + N_{Pt} \qquad (9a)$$

$$M_{Al} = N_{Al} \qquad (9b)$$

Following, the interdiffusion coefficients are estimated after modifying the Equation 4 as [37]

$$\widetilde{D}\left(Y^*_{i(N)}\right) = \frac{1}{2t}\left(\frac{dx}{dY_{MC_{i(N)}}}\right)_{Y^*_i}\left[\left(1 - Y_{MC_{i(N)}}\right)\int_{x-\infty}^{x^*}Y_{MC_{i(N)}}dx + Y_{MC_{i(N)}}\int_{x^*}^{x+\infty}\left(1 - Y_{MC_{i(N)}}\right)dx\right]$$

(10)

where $Y_{MCi} = \frac{MC_i - MC_i^-}{MC_i^+ - MC_i^-}$. The modified concentration of component $i$ (Ni or Al) is expresses as $MC_i = \frac{M_i}{V_m}$, $V_m$ is the molar volume. These values in this system are shown in Figure 17.

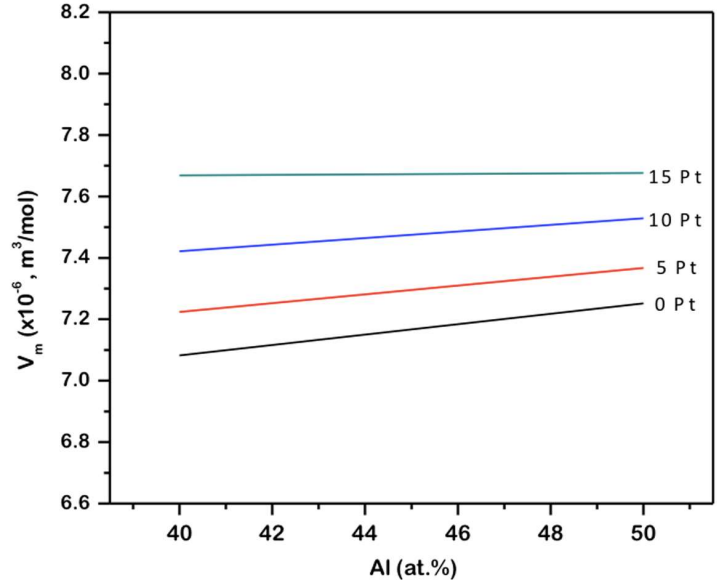

Figure 17 Molar volume variation in β-Ni(Pt)Al [27, 34],

The interdiffusion coefficients estimated in binary and pseudo-binary systems are presented in Figure 18 showing the effect pf Pt addition. The data estimated for the binary Ni-Al system are similar to the data reported by others [29, 33]. As can be seen in Figure 18, there is a significant increase in interdiffusion coefficients with Pt addition, which is more prominent close to the stoichiometry *i.e.* 50(Ni+Pt):50Al. The difference decreases with the decrease in Al content or the increase in Ni content. Intrinsic diffusion coefficients of the diffusing components $D_{Ni}$ and $D_{Al}$ are related to the the thermodynamic parameter $\emptyset = \frac{dlna_{Ni}}{dlnN_{Ni}} = \frac{dlna_{Al}}{dlnN_{Al}}$ and the tracer diffusion coefficients of the same components $D^*_{Ni}, D^*_{Al}$ by [39]

$$D_{Ni} = \frac{V_m}{\overline{V}_{Al}}D^*_{Ni}\emptyset$$

(11a)

$$D_{Al} = \frac{V_m}{\overline{V}_{Ni}}D^*_{Al}\emptyset$$

(11b)

$a_{Ni}$, $a_{Al}$ are the activities and $N_{Ni}$, $N_{Al}$ and compositions of Ni and Al. Similarly, the interdiffusion coefficient is related to the intrinsic and tracer diffusion coefficients by [37, 39]

$$\widetilde{D} = C_{Al}\overline{V}_{Al}D_{Ni} + C_{Ni}\overline{V}_{Ni}D_{Al} = (N_{Al}D^*_{Ni} + N_{Ni}D^*_{Al})\emptyset$$

(12a)

In a pseudo-binary system, the same can be expressed as [37]

$$\widetilde{D} = (M_{Al}D^*_{Ni} + M_{Ni}D^*_{Al})\emptyset$$

(12a)

Tracer diffusion coefficients measure the diffusion rates of components in the absence of the driving forces. These depend on the point defects present in a particular crystal structure. On the other hand, intrinsic diffusion coefficients are additionally influenced by the thermodynamic driving forces, which are expressed with respect to the thermodynamic factor, $\emptyset$. Therefore, we need to analyze both the thermodynamic factors and the concentration of defects for understanding the role of influencing factors on diffusion rates of components.

First, to start with the binary Ni-Al system, activities of the components in this phases are calculated theoretically [56] and measured experimentally [57]. The variation of the thermodynamic factors is shown in Figure 20 in the Ni-rich part of the β-NiAl phase, which is extracted from CALPHAD utilizing the data reported by Lu et al. [56]. In many solid solution systems, the variation of the thermodynamic factor is found to have a strong influential role in the trend of variation of the interdiffusion coefficients [58-68]. On the other hand, in the β-NiAl phase, we found that it decreases but the interdiffusion coefficient increases with the increase in Ni content. Therefore, thermodynamics driving force cannot explain the variation of the interdiffusion coefficients with composition. Rather, the trend is just opposite.

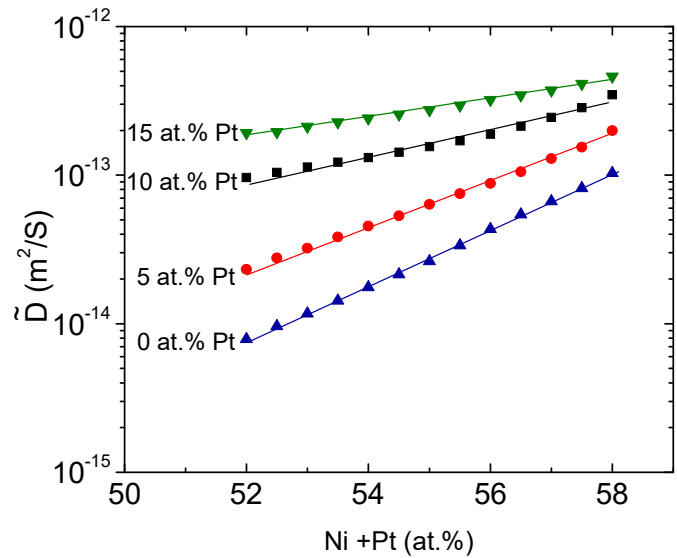

Figure 18 The variation of the interdiffusion coefficients in β-Ni(Pt)Al at 1100 °C [34, 49, 69].

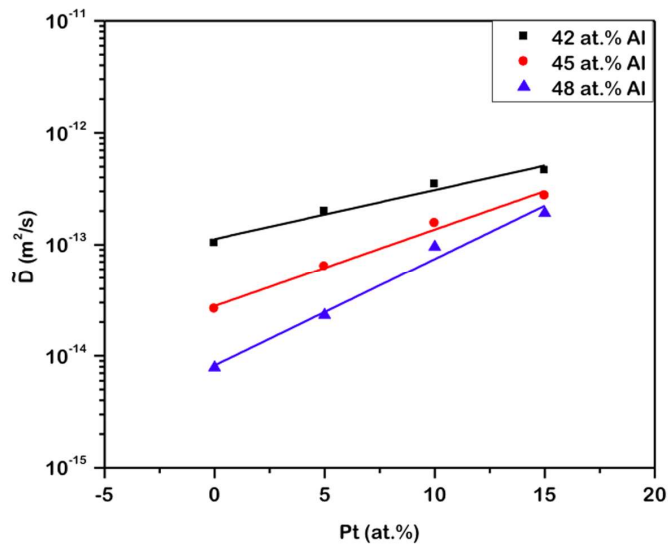

Figure 19 The variation of the interdiffusion coefficients in β-Ni(Pt)Al at 1100 °C with respect to the Pt content [34].

Therefore, it is apparent that defects play a dominant role over the thermodynamic driving forces. As already discussed, the variation of concentration of defects calculated at $0.75T_m$ by Divinski [39] is shown in Figure 12. The data are estimated at a higher temperature compared to the temperature at which the interdiffusion coefficients are estimated experimentally i.e. 1100 ºC. However, we should not expect much difference in data on defect concentration at this temperature, especially on the Ni-rich side of the phase since mainly the structural defects $Ni_{Al}$ controls the diffusion process [39]. $Ni_{Al}$ and $V_{Ni}$ are the dominant defects compared to $Al_{Ni}$ and $V_{Al}$ in the Ni-rich side. Kao and Chang [70] proposed the antistructure bridge mechanism as an atomic mechanism of diffusion on the Ni-rich side. Marino and Carter [52] studied this by first-principles density functional theory calculations and found that the diffusion rate of Ni should increase with the increase Ni content on the Ni-rich side because of increase in the concentration of Ni antisites. This was indeed found by the tracer diffusion measurements [71]. Further, they postulated that the diffusion of Al is assisted by short range next-neighbor jumps because of exchanging of positions with vacancies on Al-sublattice, which are created during diffusion of Ni. It is not possible to conduct tracer diffusion studies of Al because of the very short half-life of Al radioisotopes. On the other hand, the Kirkendall marker experiments by the diffusion couple technique indicate Ni has higher diffusion rate compared to Al on the Ni-rich side of the β-NiAl phase; however, Al also has significant diffusion rate [32, 33]. This supports the theoretical study by Marino and Carter [52].

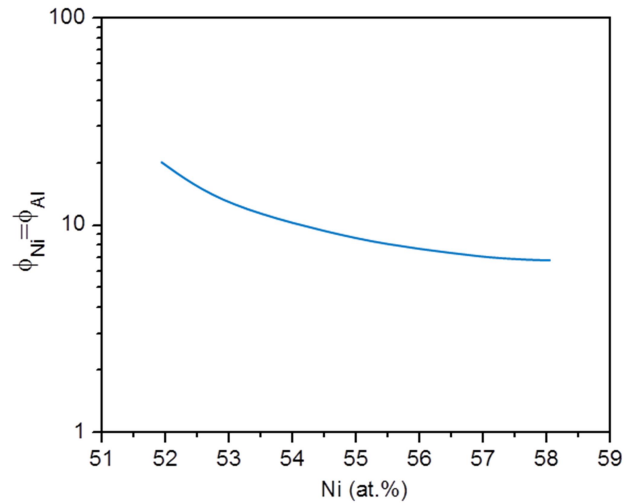

Figure 20 Variation of the thermodynamic factor in Ni-rich side of β-NiAl [49].

As already mentioned (Figure 18), the pseudo-binary experiment indicates the increase in the interdiffusion coefficient because of Pt addition in the Ni-rich side of the β-Ni(Pt)Al phase. van Loo [72] has shown that four different thermodynamic factors influence the interdiffusion coefficients in a ternary system. However, it can be simplified in our analysis. In the pseudo-binary diffusion couple, the composition of Pt is kept constant and since the diffusion coefficient of Ni is higher than the diffusion coefficient of Al in the Ni-rich side of the β-Ni(Pt)Al phase, we can examine the influence of the thermodynamic driving force by examining the influence of Pt on the thermodynamic facort for Ni $i.e.$ $\emptyset_{Ni,Ni} = \frac{dln a_{Ni}}{dln x_{Ni}}$. The composition dependent variation of this parameter extracted in CALPHAD using the data published by Lu et al. [56] and is shown in Figure 21. The trend of variation of the thermodynamic factor with composition remains the same even after the addition of Pt. Additionally, it decreases with the increase in Pt content. Therefore, defects must have a dominating role over the thermodynamic factor. Marino and Carter [52] calculated the defect formation and migration energies for diffusion of Ni and Al in the presence of Pt. The values are given in Table 1. It can be seen that these values decrease because of Pt addition justifying the increase in diffusion rate of Ni and Al leading to increase in interdiffusion coefficients. However, a dedicated study is required to correlate the experimental results with extensive theoretical calculations. Another study conducted previously has shown that the growth rates of both γ-NiAl and γ'-Ni₃Al phases increase with the addition of Pt indicating a similar effect of Pt in these phases [55]. As shown in Figure 12 and 18, the interdiffusion coefficients increase with the increase in Ni content because of increase in concentration of Ni antisites. Ni tracer diffusion studies [33, 71] indicate the increase in diffusion coefficient with the increase in Ni content. Another important fact should be noted in Figure 18 and 19. The effect of Pt addition on the increase in interdiffusion coefficients is less prominent in the Ni-rich side. Therefore, it is well possible that some of the Ni antisites are occupied by Pt since Pt occupies the same sublattice as Ni.

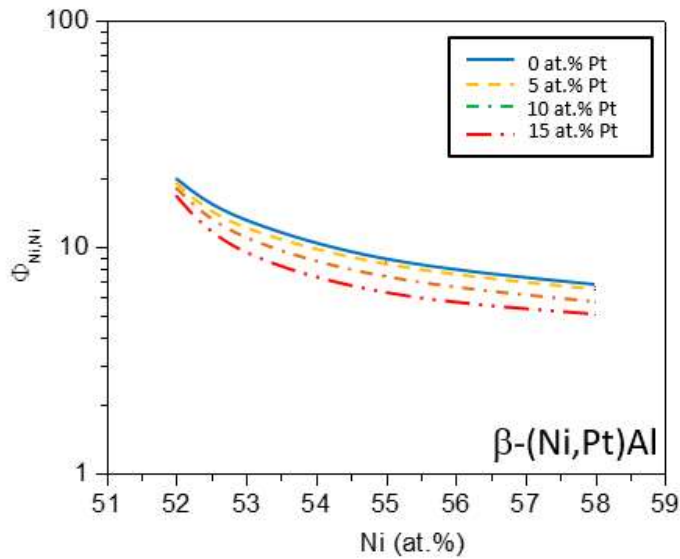

Figure 21 The variation of the thermodynamic factor with respect to the Ni and Pt content in β-Ni(Pt)Al

Table 1a Ni diffusion in Ni-rich NiAl: Defect formation energy, migration energy and activation energy in [100] direction for 0 Pt and 1.9 at. % Pt [52].

| Pt content | Defect formation energy, $E_f$ (eV) | Migration energy, $E_m$ (eV) | Activation energy, Q (eV) |
|---|---|---|---|
| 0 Pt | 1.67 | 0.67 | 2.34 |
| 1.9 at.% Pt | 1.06 | 0.65 | 1.71 |

Table 1b Al diffusion in Ni-rich NiAl: Defect formation energy, migration energy and activation energy in [100] direction for 0 Pt and 1.9 at. % Pt [52].

| Pt content | Defect formation energy, $E_f$ (eV) | Migration energy, $E_m$ (eV) | Activation energy, Q (eV) |
|---|---|---|---|
| 0 Pt | 2.02 | 1.24 | 3.26 |
| 1.9 at.% Pt | 1.43 | 1.04 | 2.47 |

## 2.1.3 Effect of Pt on growth kinetics and microstructural evolution of IDZ between bond coat and René N5

In this section, we report the effect of Pt addition in a bond coat on growth kinetics of the interdiffusion zone between bond coat and superalloys with the help of fundamental diffusion studies discussed in the previous sections. Two different sets of bond coat alloys were prepared by arc melting which is designated as

Type 1: $(Ni_{50-x}Pt_x)Al_{50}$

Type 2: $(Ni_{60-x}Pt_x)Al_{40}$

where x = 0, 5, 10 and 15 at.%. Therefore, one can understand from Figure 22 that the type 1 bond coat has an average composition of (Ni+Pt): Al≡50:50 and the type 2 bond coats have an average composition of (Ni+Pt): Al≡60:40. Type 2 bond coat is close to the (Ni,Pt)-rich phase boundary composition. Two different superalloys René N5 and CMSX-4 are used in this study to couple with the bond coats. These are the phase mixtures of γ-Ni(Al) solid solution and γ´-Ni₃Al intermetallic compound. Many refractory components are alloyed in both the superalloys as listed in Table 2 which are partitioned differently in these phases. It can be seen that the difference in composition in these two superalloys is very less. Ti is present in CMX4 and it not there in René N5.

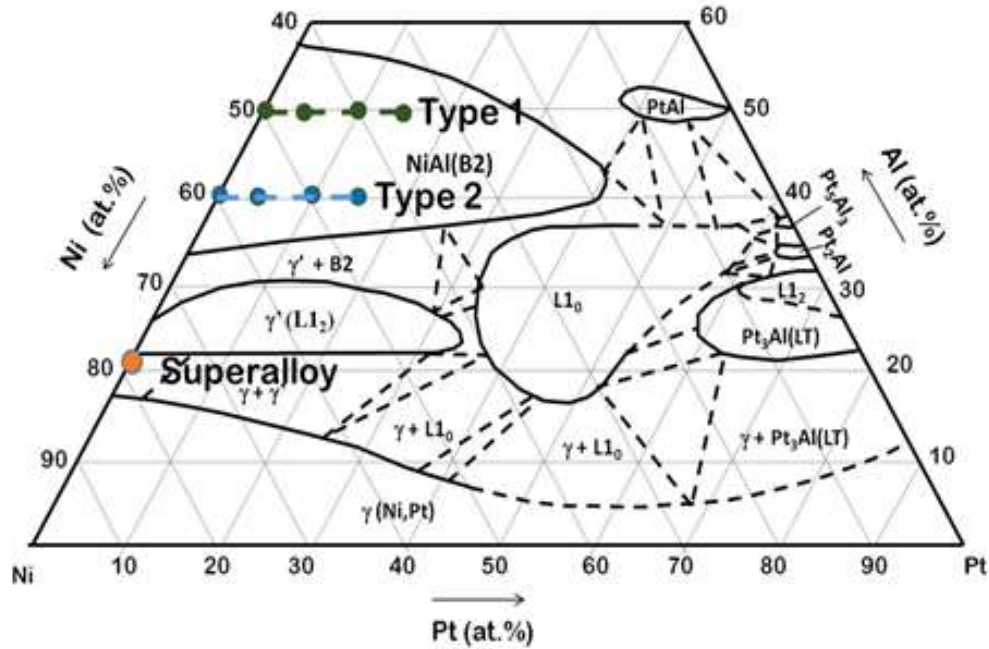

Figure 22 Different types of bond coats and the superalloys used in these study are shown on a Ni-Pt-Al phase diagram (phase diagram is adapted from [26]).

Table 2 The average compositions of two different superalloys used in this study are listed.

| Element (at.%) | Co | Cr | Al | Ti | Ta | Mo | W | Re | Ni |
|---|---|---|---|---|---|---|---|---|---|
| René N5 | 7.7 | 8.1 | 13.8 | - | 2.2 | 0.94 | 1.6 | 1 | 64.7 |
| CMSX-4 | 8.7 | 7.1 | 11.8 | 1.2 | 2 | 0.3 | 1.8 | 1 | 66 |

ppm levels: Y, Hf, Zr, S, C, O, P

Figure 23 shows an interdiffusion zone between type 1 bond coat with 15 at.% Pt and René N5 annealed at 1100 °C for 25 h. It has mainly two parts separated by the Kirkendall marker plane;

precipitate free interdiffusion zone which could be detected by the composition profile and precipitate containing interdiffusion zone, which can be easily seen in the SEM micrograph because of the presence of precipitates with bright contrast. The location of the Kirkendall marker plane (indicated as K) is detected by the presence of a line of pores [39]. This is shown in an enlarged SEM micrograph in Figure 24. The location of this micrograph in Figure 23 is shown by a dotted rectangle. One can understand the diffusion-controlled growth and microstructural evolution following the physico-chemical model by the present author [42, 43]. This is successfully used in many other systems to quantify the diffusion parameters without using inert particles as the Kirkendall markers [39]. It should be noted here that the location of this plane before diffusion annealing of the diffusion couple of two dissimilar materials is at the initial bonding interface. It is relocated after the diffusion annealing depending on the relative mobilities of the components. In the present example, the location of this plane indicates that the precipitate free zone is grown from the bond coat and the precipitate containing zone grows from the superalloy.

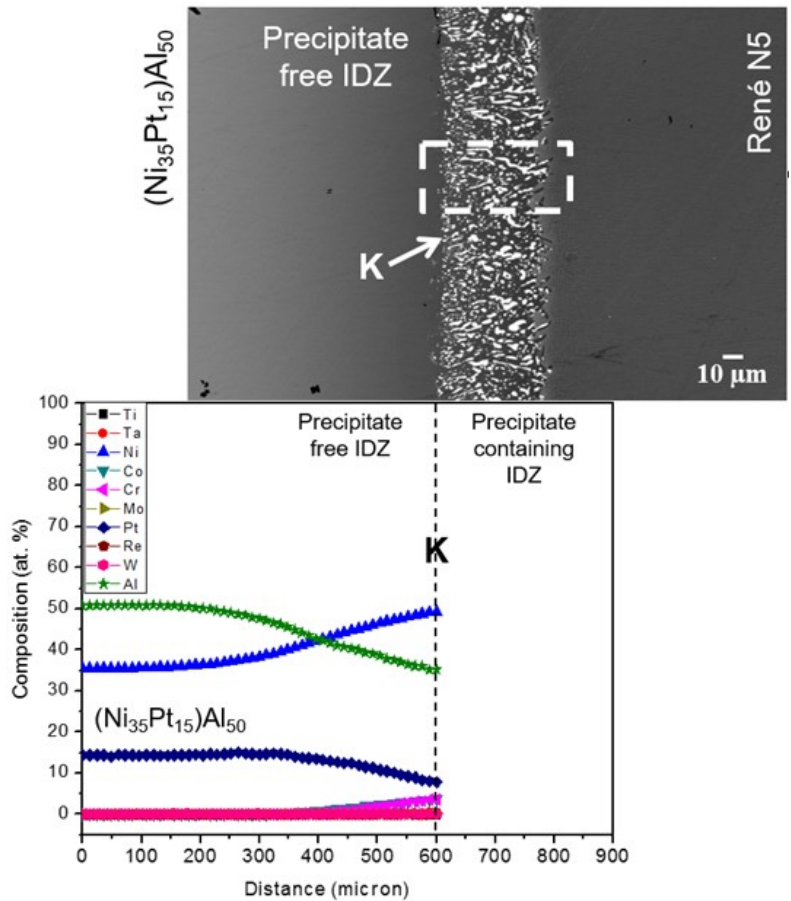

Figure 23 Interdiffusion zone between type 1 β-(Ni$_{35}$,Pt$_{15}$)Al$_{50}$ and René N5 annealed at 1100 °C for 25 h [49].

The composition profile in Figure 23 indicates that the precipitate-free IDZ grows from the bond coat by the of loss of Pt and Al from the bond coat because of composition difference between the bond coat and the superalloy. On the other side of the marker plane, as can be seen in Figure 24, the loss of Ni leads to the growth of precipitates. In a superalloy, refractory components are used in very high concentration maximizing the amount that can be dissolved in γ-Ni(Al) solid solution and γ'-Ni$_3$Al. With the loss of Ni, β-phase grows in the interdiffusion zone and the precipitates are grown since the refractory components cannot be remained dissolved in this phase with such a high concentration. In between these two sublayers, a thin layer of β and γ' is found in the interdiffusion zone. This can be interpreted as γ+ γ' transform to β+ γ' by the loss of Ni and with further loss, it transforms to β.

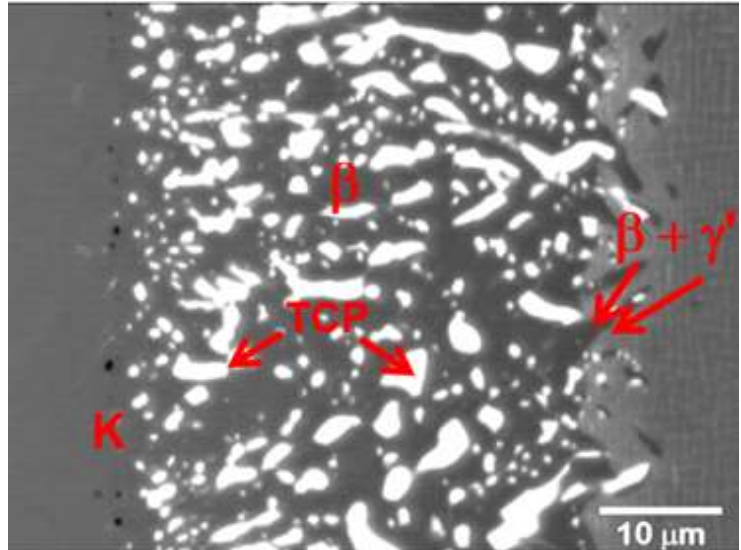

Figure 24 A focused area of the interdiffusion zone, as shown in Figure 23, is highlighted [49].

As already mentioned, the composition of type 2 bond coats are located at very close to the Ni(Pt)-rich phase boundary of the β-Ni(Pt)Al phase, as shown in Figure 22. As a result, the precipitate free interdiffusion zone does not grow with a large variation of compositions prominently when type 2 bond coats are coupled with the superalloys. This can be realized from the composition profile, as shown in Figure 25. The precipitate containing IDZ looks similar but has lower thickness compared to type 1 bond coat. A schematic representation of the growth process is shown in Figure 26.

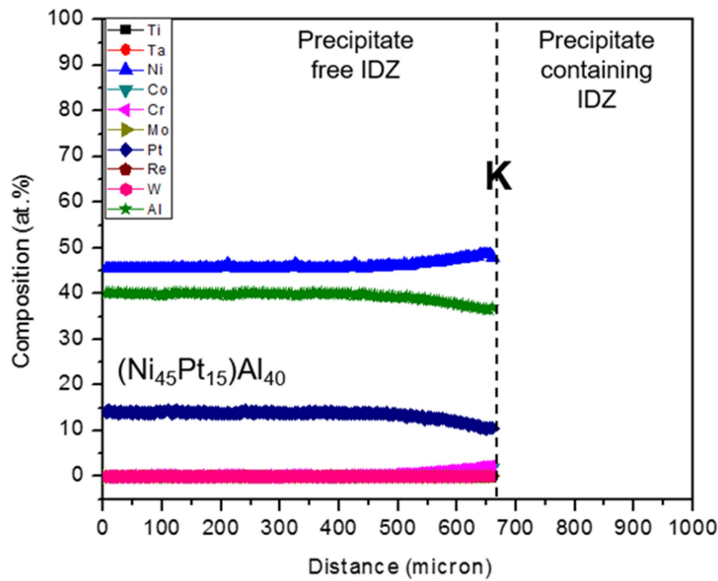

Figure 25 Composition profile of precipitate free interdiffusion zone between type 2 bond coat and René N5 annealed at 1100 °C for 25 h [49].

Figure 27 compares the effect of Ni (or Al) and Pt content in the bond coat on the growth of precipitate containing interdiffusion zone. Two examples of 0 and 15 at.% Pt are shown for both types of bond coats are shown when coupled with René N5. In a SEM precipitate, only the precipitate containing zone can be seen clearly; whereas, the precipitate free part can be realized from a composition profile. The variations of both the parts for the different content of Pt are listed in Table 3. It can be found easily that the thicknesses of both the sublayers increase significantly with the increase in Pt content. Previously, we have shown that diffusion rates of components increase in the presence of Pt and therefore it is apparent that the diffusion controlled growth of the interdiffusion zone also increases with the addition of Pt.

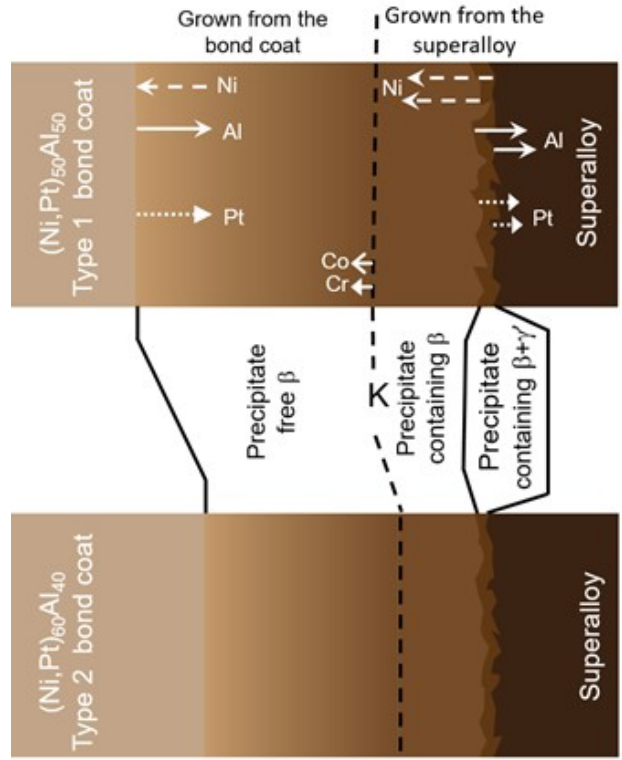

Figure 26 A schematic representation of the growth of the interdiffusion zone between the bond coats and the superalloy [49].

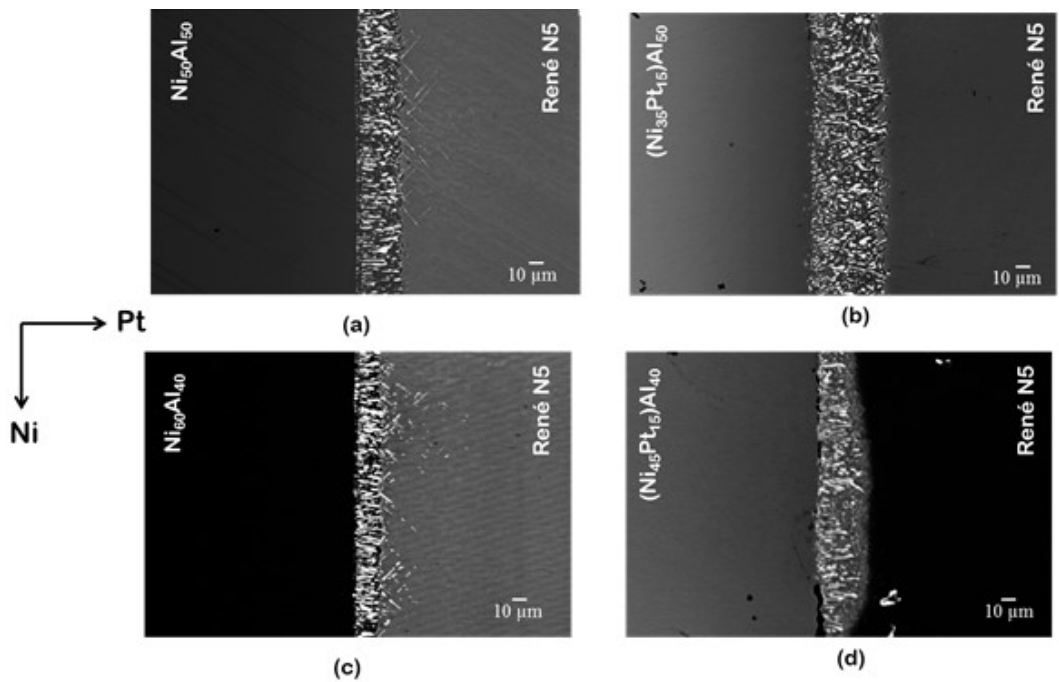

Figure 27 The comparison of growth of the interdiffusion zone because of Ni (or Al) and Pt variation in β-Ni(Pt)Al and René N5 diffusion couple annealed at 1100 °C for 25 h [49]

Table 3 Thicknesses of sublayers in the interdiffusion zone between bond coats/René N5 [49]

| Bond coat composition (type 1) | Thickness of precipitate free IDZ (µm) | Thickness of precipitate containing IDZ (µm) | Bond coat composition (type 2) | Thickness of precipitate containing IDZ (µm) |
|---|---|---|---|---|
| $(Ni_{50})Al_{50}$ | $356 \pm 7$ | $21 \pm 0.9$ | $(Ni_{60})Al_{40}$ | $15 \pm 0.6$ |
| $(Ni_{45}Pt_5)Al_{50}$ | $368 \pm 7$ | $26 \pm 1.3$ | $(Ni_{55}Pt_5)Al_{40}$ | $21 \pm 1$ |
| $(Ni_{40}Pt_{10})Al_{50}$ | $520 \pm 10$ | $34 \pm 1.7$ | $(Ni_{50}Pt_{10})Al_{40}$ | $22 \pm 1.1$ |
| $(Ni_{35}Pt_{15})Al_{50}$ | $570 \pm 11$ | $40 \pm 2$ | $(Ni_{45}Pt_{15})Al_{40}$ | $30 \pm 1.5$ |

Another important aspect should be noted in Table 3 that the thickness of precipitate-containing IDZ is higher for type 1 bond coat compared to type 2 bond coat. As already discussed before, the diffusion profile of precipitate free IDZ is more prominent for type 1 compared to type 2 bond coat. Therefore, the loss of Ni from the superalloy is also higher leading to higher rate of growth of precipitates in the case of type 1 bond coat. Further, as we have shown that the addition of Pt increases the diffusion controlled growth rate of the interdiffusion zone because of increase in diffusion coefficients of the components. Therefore, as expected, the thickness of precipitate containing interdiffusion zone increases with the faster rate in the case of type 1 compared to type 2 bond coat. The growth of TCP phase is deleterious to the structure. Therefore, this study indicates that although the addition of Pt improves the oxidation resistance it increases the growth rate of the precipitates. Therefore, the content of Pt should be optimized to find a property balance.

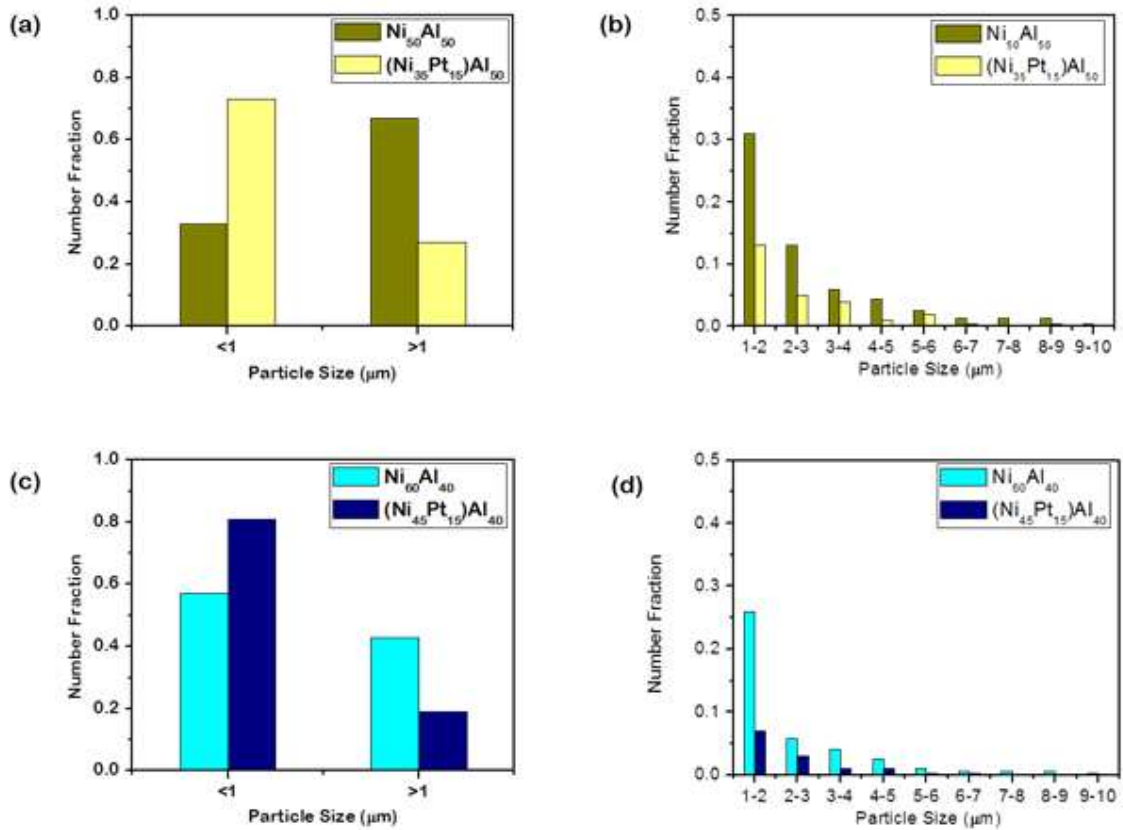

Figure 28 Effect of Pt additions in type 1 and type 2 bond coats on precipitate size distribution [49].

There is an effect of Pt addition on the size distribution of precipitates. The precipitates are, in general, relatively long. Few precipitates are long enough to cover the thickness of the interdiffusion zone containing precipitates. As already mentioned, the interdiffusion zone thickness increases with the increase in Pt addition. However, the area fractions of precipitates remain more or less the same. Therefore, the total amount of precipitate grows with the addition of Pt. To get a better understanding of the size distribution, the sizes of the precipitates are first measured in two groups, less and higher than 1 μm. 700-800 precipitates are measured for this distribution. As shown in Figure 28, the number of smaller precipitates increases drastically with the addition of Pt in both types of bond coats. Moreover, the number of precipitates of bigger sizes decreases. Therefore, this study indicates that the addition of Pt enhances the nucleation that leading to decrease in the average size of precipitates.

To get an overall idea of the precipitate type and composition distribution, WDS mapping of the interdiffusion zone containing precipitate is conducted in electron probe microanalyzer (EPMA), as shown in Figure 29 for $(Ni_{35}Pt_{15})Al_{50}$/René N5 couple. Very high "Z" contrast indicates that the precipitates are rich with heavy refractory components. Based on composition distribution of Co and Cr, it is very difficult to identify if there are different types of precipitates present. However, when we compare the Ta and W distribution, it is clear that there must be two types of precipitates (at least). Ta and W-rich precipitates are bigger in size.

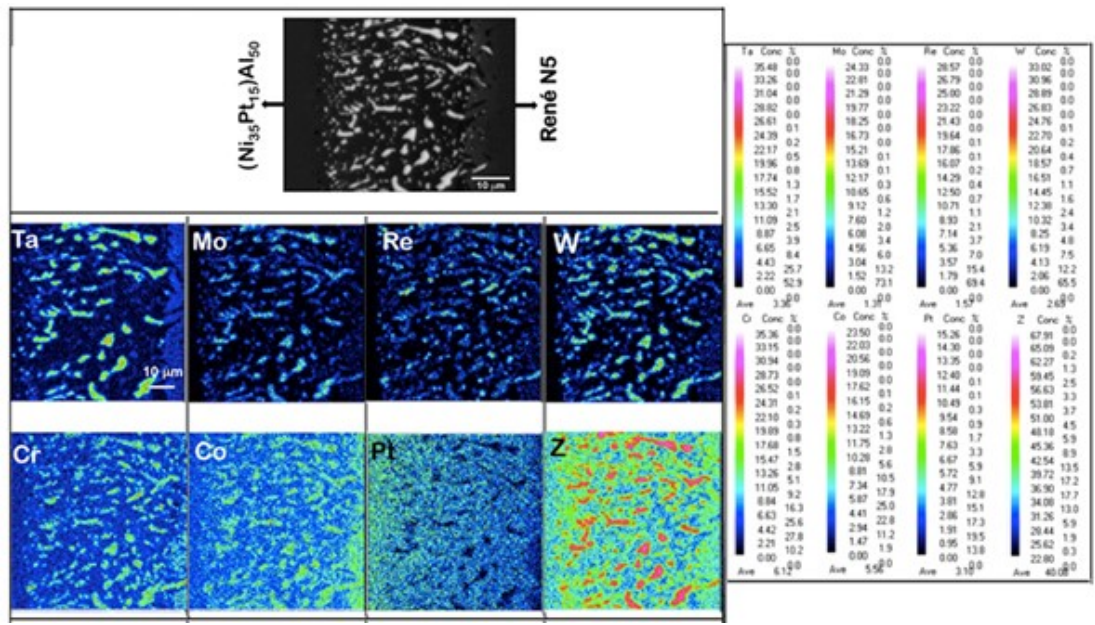

Figure 29 Elemental distribution in the precipitate containing interdiffusion zone of $(Ni_{35}Pt_{15})Al_{50}$/René N5 diffusion couple [49].

Transmission electron microscopy is used to analyze further on the precipitate types. Figure 30 presents a bright field image showing few precipitates. At few places, few types of precipitates are found as attached together with a head and tail type feature. An energy dispersive spectroscopy line profile measurement covering the whole length of two bonded precipitates indicate the presence of Ta and W rich and lean precipitates. The head part has a higher concentration of Ta and W compared to the tail part. Diffraction patterns confirm the types of precipitates. The head part is indexed as the μ phase (rhombohedral) and the tail part is indexed as the P phase (orthorhombic). Rae and Reed [56] found the presence of the same precipitates in superalloys.

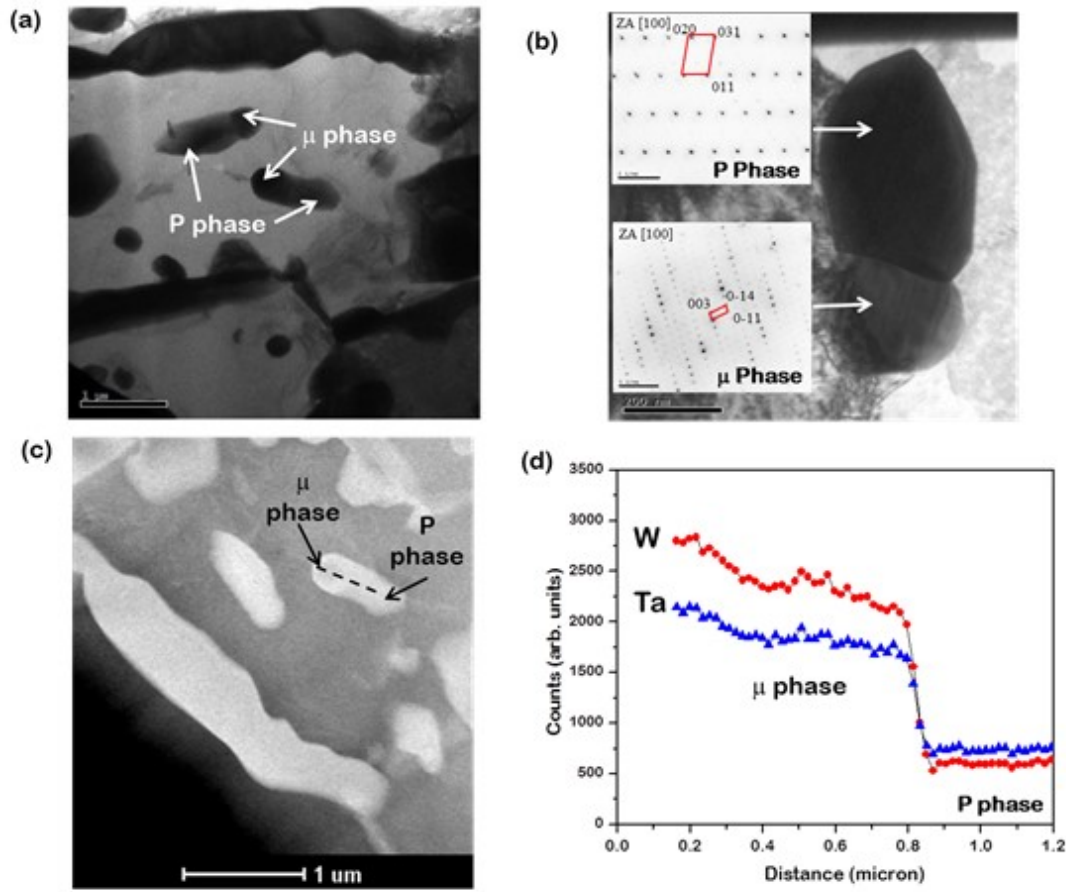

Figure 30 TEM analysis of precipitates in $(Ni_{35}Pt_{15})Al_{50}$/René N5 diffusion couple [49].

## 2.1.4 Effect of Pt on growth kinetics and microstructural evolution of IDZ between bond coat and CNX4

A similar line of experiments was conducted with another type of superalloy CMX4. The interdiffusion zone is shown in Figure 31. With respect to the general features, the interdiffusion zone is similar to René N5. It has two parts, precipitate free and precipitate containing interdiffusion zones. Precipitate free part can be realized from the measured composition profile. On the other hand, precipitate containing part can be seen in the figure. An enlarged part, as marked in Figure 32a, is shown in Figure 32b. It is apparent that with the loss of Ni, the γ'-Ni₃Al+ γ'-Ni(Al) phase mixture of superalloy first transform to β-NiAl + γ'-Ni₃Al and then convert to β. Precipitates are grown in the β-phase with high volume fractions because of low solubility limit of the refractory components in the β - phase compared to the γ and γ' phases. The variations of both the sublayers are listed in Table 4. Similar to the results of bond coat/René N5 system, the precipitate containing interdiffusion zone thickness higher for type 1 compared to the type 2 bond coat. Precipitates are also found further inside the superalloy. It should be noted here that precipitates are also found inside the superalloys occasionally, which are grown during thermal aging because of thermal and thermodynamic instability as reported earlier by Rae and Reed [74].

As already discussed above, there is not much difference in growth behavior of the interdiffusion zone when two different types of superalloys are used. However, a clear difference in the size of precipitates can be seen. Precipitates in the case of CMX4 are much finer compared to René N5.  As shown in Figure 32, in this case also two types of precipitates are found $\mu$ (rhombohedral) and $\sigma$ (tetragonal), as confirmed by the diffraction pattern analysis in TEM.

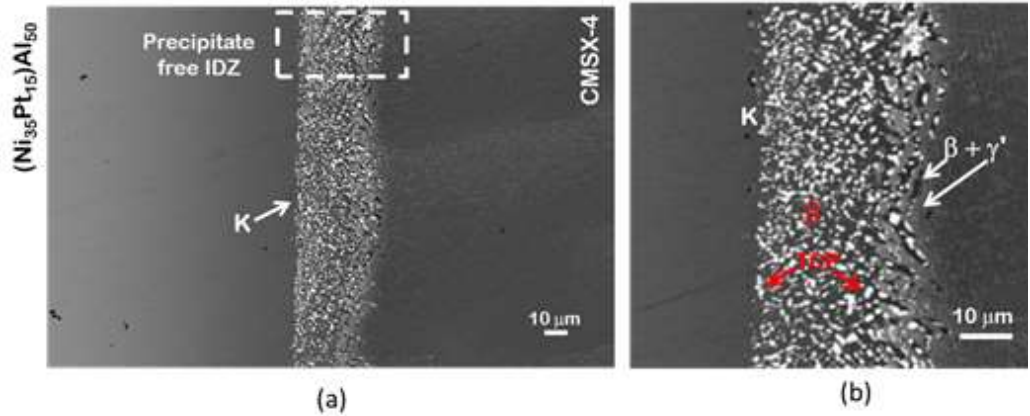

Figure 31 Interdiffusion zone between type 1 $\beta$-(Ni$_{35}$,Pt$_{15}$)Al$_{50}$ and CMX4 annealed at 1100 °C for 25 h [49].

Table 4 Thicknesses of sublayers in the interdiffusion zone between bond coats/CMX-4 [49].

| Bond coat composition (type 1) | Thickness of precipitate free IDZ (μm) | Thickness of precipitate containing IDZ (μm) | Bond coat composition (type 2) | Thickness of precipitate containing IDZ (μm) |
|---|---|---|---|---|
| (Ni$_{50}$)Al$_{50}$ | 230 ± 4 | 16 ± 0.8 | (Ni$_{60}$)Al$_{40}$ | 10 ± 0.5 |
| (Ni$_{45}$Pt$_5$)Al$_{50}$ | 245 ± 5 | 21 ± 1 | (Ni$_{55}$Pt$_5$)Al$_{40}$ | 12 ± 0.6 |
| (Ni$_{40}$Pt$_{10}$)Al$_{50}$ | 307 ± 6 | 31 ± 1.6 | (Ni$_{50}$Pt$_{10}$)Al$_{40}$ | 18 ± 0.9 |
| (Ni$_{35}$Pt$_{15}$)Al$_{50}$ | 500 ± 10 | 35 ± 1.8 | (Ni$_{45}$Pt$_{15}$)Al$_{40}$ | 25 ± 1.5 |

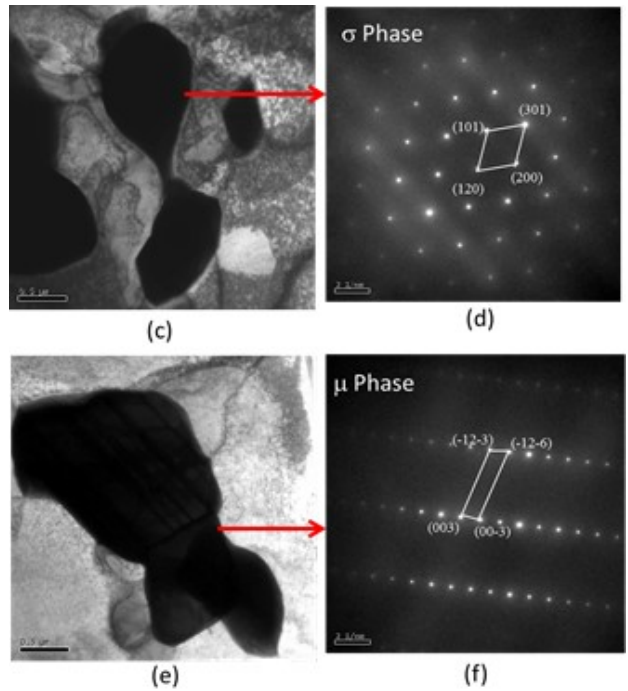

Figure 32 TEM analysis of precipitates in (Ni$_{35}$Pt$_{15}$)Al$_{50}$/CMX 4 diffusion couple [49].

Figure 33 shows the effect of Pt addition in both types of bond coat alloys when coupled with CMX4. It can be seen that the thickness of the interdiffusion zone increases because of increase in diffusion coefficients of components in the presence of Pt. The thickness values are listed in Table 4.

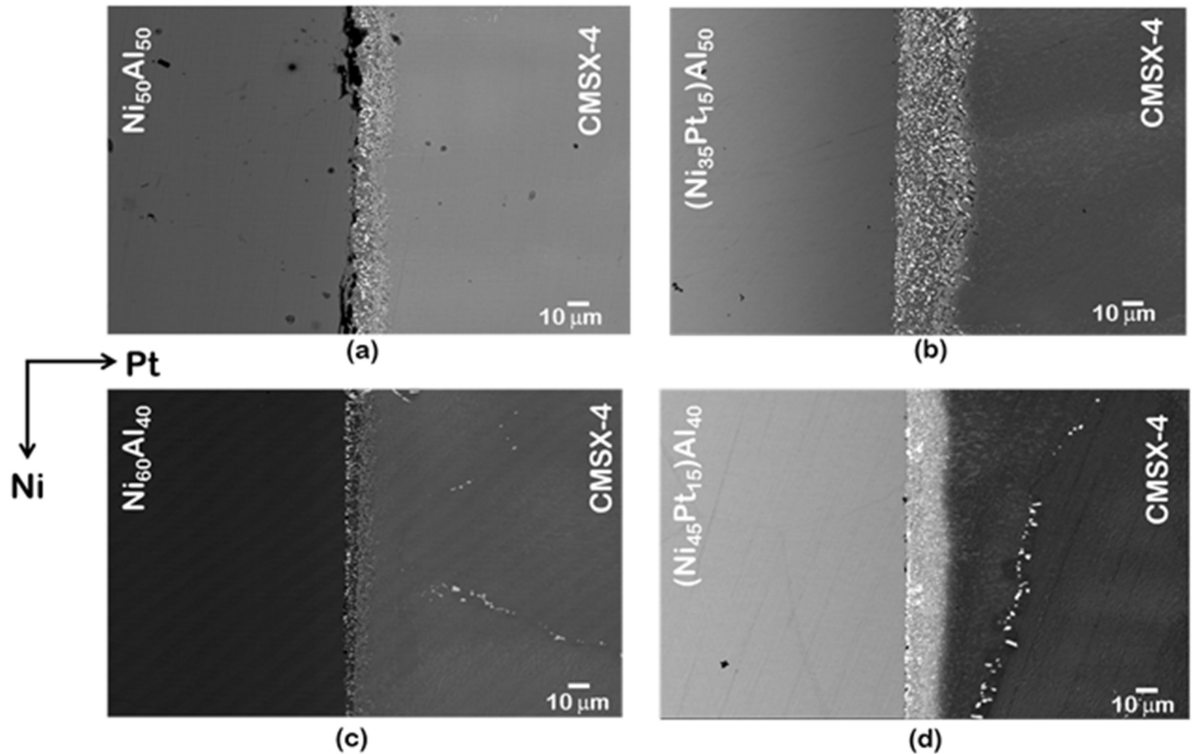

Figure 33 The comparison of growth of the interdiffusion zone because of Ni (or Al) and Pt variation in β-Ni(Pt)Al and CMX4 diffusion couple annealed at 1100 °C for 25 h [49]

## 2.2 Microstructural evolution between steel and aluminide coating

Hot dip aluminization of steel is one of the favorable techniques to coat steels for the use in high-temperature applications, which otherwise cannot withstand such a harsh environment. The formation of $Al_2O_3$ provides the necessary oxidation resistance. There are few studies available on the growth of different aspects of the product phases during hot dip aluminization [75-79]. As shown in Figure 34, the interdiffusion zone mainly consists of the $Fe_2Al_5$ phase with a tongue like microstructural features [75]. In between Al and $Fe_2Al_5$, a very thin layer of $FeAl_3$ is found. At the steel/$Fe_2Al_5$ interface two other intermetallic compounds, $Fe_3Al$ and $FeAl$ are found as very thin layers. The irregular shape of $Fe_2Al_5$ is still under investigation. The reaction diffusion process in the liquid state must have a role; however, the role of anisotropic diffusion cannot be ruled out [75]. The waviness in the growth of the product phase is commonly found during soldering with a Sn-based alloy on Cu [80]. On the other hand, such a tongue like the feature of the product phase is found even during solid-state reaction diffusion between Ni and Si [81]. This is related to the anisotropic diffusion in this system.

Bhanumurthy et al. [82] conducted solid-state diffusion controlled the growth of the phases following the bulk diffusion couple technique in the temperature range of 450-600 °C. Similar to the steel/aluminium reaction diffusion, mainly the $Fe_2Al_5$ is found to grow in the interdiffusion zone. Other phases, such as $Fe_3Al$, $FeAl$, and $FeAl_2$ are found to grow only at or above 600 °C. The activation energy for the interdiffusion in the $Fe_2Al_5$ is found to be 146.8 kJ/mol.

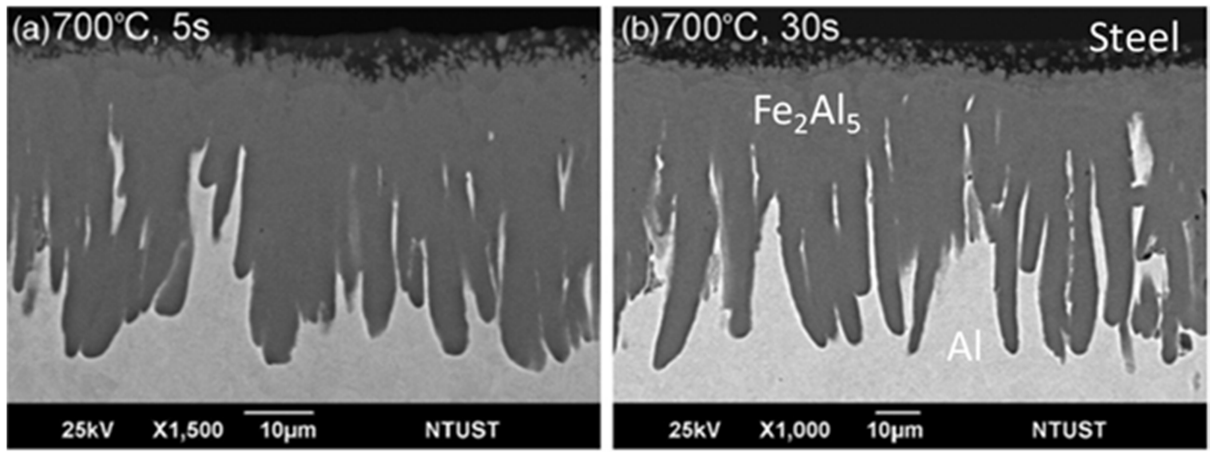

Figure 34 Hot dip aluminization of mild steel for different times [75].

Pint et al. [83] developed a Fe-aluminide coating with superior composition control by the chemical vapor deposition process, as shown in Figure 35. At the outer surface, Al-rich thin layer and below that relatively Al-lean layer is grown. As already mentioned, the number of studies in these systems are not adequate and further dedicated studies are required to develop a better understanding on diffusion-controlled growth-mechanism of the phases.

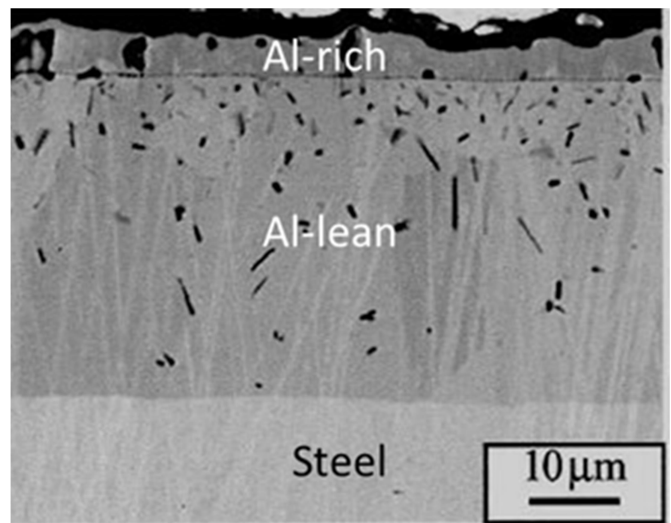

Figure 35 The Fe-aluminide coating on steel developed by chemical vapor deposition [83].

## 3. Conclusion

We can summarize different aspects as:

(i) Quantitative diffusion analysis is conducted in the β-NiAl phase to study the influence of Pt on interdiffusion coefficients of Ni and Al following a pseudo-binary approach. It is found that the addition Pt increases the diffusion rates of components. The defects present in the structure are found to play a dominating role over the thermodynamic driving forces.

(ii) The interdiffusion zone between the bond coat and the superalloys are studied by coupling two types of bond coats with two different superalloys (René N5 and CMX4). It grows with two parts. A precipitate free interdiffusion zone grows from the bond coat mainly by the loss of Al and Pt. A precipitate containing interdiffusion zone grows from the superalloy mainly by the loss of Ni.

(iii)    The growth of the precipitates with the Ni-rich bond coat is found to be lower, which is beneficial with respect to the reliability of the structure since precipitates are brittle and therefore deleterious to the structure.

(iv)    Pt addition increases the growth rate of the interdiffusion zone. This is expected since diffusion rates of the components increase with the Pt addition. Because of higher diffusion of Ni, the growth rate of the precipitates also increases. With the loss of Ni, refractory components are rejected from the substrate for the growth of these precipitates.

(v)    With respect to the general features, the interdiffusion zone is similar to René N5 and CMSX-4. However, the precipitates are much finer in the case of CMX4. Ti-free μ-phase grows in the case of both the superalloys. Additionally, P-phase grows in the case of René N5 and σ-phase grows in the case of CMX4.

(vi)    The hot dip aluminization process is used for the growth of a coating layer on steel. In this, the interdiffusion zone is covered mainly by $Fe_2Al_5$ phase. Other phases are present with the very thin layer. At the same time, CVD process is being developed for a better control of the coating layer.

## References

[1] N.P. Padture, M. Gell, E.H. Jordan, Science 296 (5566), 280-284.

[2] https://www.phase-trans.msm.cam.ac.uk/2003/Superalloys/coatings/images/compare_coatings.jpg

[3] D.K. Das, Prog. Mater. Sci. 58 (2013) p. 151-182.

[4] G.R. Krishna, D.K. Das and S.V. Joshi, Mater. Sci. Eng. A 251 (1998) 40-47.

[5] P. Deb, D.H. Boone, R. Streiff, R.D. Sisson Jr. (Ed.), Surface modifications and coatings, ASM, Metals Park, Ohio (1986) 143–159.

[6] R. Streiff, O. Cerclier, D.H. Boone, Surf. Coat .Technol. 32 (1987) 111–126.

[7] H.M. Twancy, N. Sridhar, B.S. Tawabini, N.M. Abbas, T.N. Rhys-Jones, J. Mater. Sci. 27 (1992) 6463–6474.

[8] G.J. Tatlock, T.J. Hurd, J.S. Punni, Platin. Met. Rev. 31 (1987) 26–31.

[9] J.G. Fountain, F.A. Golightly, F.H. Scott, G.C. Wood, Oxid. Met. 10 (1976) 341–345.

[10] G.R. Johnston, J.L. Cocking, W.C. Johnson, Oxid. Met. 23 (1985) 237–249.

[11] J.H.W. De Wit, P.A. Van Manen, Mater. Sci. Forum. 154 (1994) 109–118.

[12] J. Angenete, K. Stiller, Surf. Coat. Technol. 150 (2002) 107–118.

[13] D.K. Das, V. Singh, S.V. Joshi, Oxid. Met. 57 (2002) 245–266.

[14] D.K. Das, M. Roy, V. Singh, S.V. Joshi, Mater. Sci. Technol. 15 (1999) 1199–1208

[15] J. Bouhanek, O.A. Adesanya, F.H. Scott, P. Skeldon, D.G. Less, Mater. High Temp. 17 (2000) 185–196.

[16] M.W. Chen, R.T. Ott, T.C. Hufnagel, P.K. Wright, K.G. Hemkar, Surf. Coat. Technol. 163–164 (2003) 25–30.

[17] J.L. Smialek, R.F. Hehemann, Metall. Trans. 4 (1973) 1571–1575.

[18] M.W. Chen, M.L. Glynn, R.T. Ott, T.C. Hufnagel, K.J. Hemker, Acta Mater. 51 (2003) 4279–4294.

[19] V.K. Tolpygo, D.R. Clarke, Acta Mater., 48 (2000) p. 3283–3293.

[20] R. Panat, S. Zhang, K.J. Hsia, Acta Mater. 51 (2003) p. 239–249.

[21] V.K. Tolpygo, D.R. Clarke, Acta Mater. 52 (2004) p. 5115–5127.

[26] S. Hayashi, S.I. Ford, D.J. Young, D.J. Sordelet, M.F. Besser, and B. Gleeson, Acta Mater. *53* (2005) 3319–3328.

[27] N. Jaya, PhD Thesis, Micro-scale fracture testing of graded (Pt,Ni)Al bond coats, Indian Institute of Science, Bangalore, India, 2013.

[28] S. Shankar, L.L. and Seigle, Met. Trans. A, 9A (1978) 1467-76.

[29] M. Watanabe, Z. Horita and M. Nemoto, Defect Diffus. Forum 143-147 (1997) 345-50.

[30] T. Helander and J. Ågren, Acta Mater. 47 (1999) 1141-52 4.

[31] S. Kim and Y.A. Chang, Metall. Mater. Trans. A, 31A (2000) 1519-24

[32] A Paul, AA Kodentsov, FJJ Van Loo, Acta materialia 52 (2004) 4041-4048.

[33] A. Paul, A.A. Kodentsov and F.J.J. van Loo, J. Alloys and Compd. 403 (2005) 147-153.

[34] P Kiruthika, A Paul, Philos. Mag. Lett. 95 (2015) 138-144.

[35] A. Paul, Philos. Mag., 93 (2013) 2297-2315.

[36] S Santra, A Paul, Scripta Materialia 103 (2015) 18-21.

[37] A Paul, Scripta Mater. 135 (2017) 153-157.

[38] A. Paul, The Kirkendall effect in solid state diffusion, PhD thesis, Technical University of Eindhoven, 2004.

[39] A Paul, T Laurila, V Vuorinen, SV Divinski, Thermodynamics, Diffusion and the Kirkendall Effect in Solids, Springer, Heidelberg, Germany, 2014.

[40] A Paul, MJH Van Dal, AA Kodentsov, FJJ Van Loo, Acta Mater. 52 (2004) 623-630.

[41] A Paul, AA Kodentsov, FJJ Van Loo, Intermetallics 14 (2006) 1428-1432.

[42] C. Ghosh and A. Paul, Acta Mater. 55 (2007) 1927-1939.

[43] C Ghosh, A Paul, Intermetallics 16 (2008) 955-961.

[44] C. Ghosh and A.Paul, Acta Mater. 57 (2009) 493-502.

[45] A Paul, J. Mater. Sci. Mater. Electron. 22 (2011) 833-837.

[46] S Santra, A Paul, Intermetallics 70 (2016) 1-6.

[47] VA Baheti, S Kashyap, P Kumar, K Chattopadhyay, A Paul, Philos. Mag. 97 (2017) 1782-1802.

[48] A. Paul, Estimation of diffusion coefficients in binary and pseudo-binary bulk diffusion couples, A Paul, S Divinski (Editors), Handbook of solid state diffusion; Diffusion fundamentals and techniques, Volume 1 (2017) 77-199.

[49] P Kiruthika, SK Makineni, C Srivastava, K Chattopadhyay, A Paul, Acta Materialia 105 (2016) 438-448.

[50] A. Paul, T. Laurila and S. Divinski, Defects, driving forces and definitions of diffusion coefficients in solids, A Paul, S Divinski (Editors), Handbook of solid state diffusion; Diffusion fundamentals and techniques, Volume 1 (2017) 1-53.

[51] S. Divinski, Defects and diffusion in ordered compounds, A Paul, S Divinski (Editors), Handbook of solid state diffusion; Diffusion fundamentals and techniques, Volume 1 (2017) 449-513.

[52] K.A. Marino and E.A. Carter, Intermetallics 18 (2010) 1470-1479.

[53] Y. Minamino, Y. Koizumi, N. Tsuji, M. Morioka, K. Hiraoand Y Shirai, Sci. Technol. Adv. Mater. 1 (2000) 237-249.

[54] VD Divya, U Ramamurty, A Paul, Philos. Mag. 93 (2013) 2190-2206.

[55] K.Y. Tsai, M.H. Tsai and J.W. Yeh, Acta Materialia 61 (2013) 4887-4897.

[56] X. G. Lu, B. Sundman and J. Ågren, CALPHAD 33 (2009) 450-456.

[57] A. Steiner and K.L. Komarek, Trans. Met. Soc. AIME, 230 (1964) 786-90.

[58] B. Mishra, P. Kiruthika and A. Paul, J. Mater. Sci. Mater. Electron. 25 (2014) 1778-1782.

[59] S. Santra, H. Dong, T. Laurila, A. Paul, Proc. R. Soc. A 470 (2013) 20130464. (DOI: 10.1098/rspa.2013.0464).

[60] V.A. Baheti, R. Ravi and A. Paul, J. Mater. Sci. Mater. Electron. 24 (2013) 2833-2838.

[61] R. Ravi and A. Paul, Diffusion mechanism in the gold-copper system, J. Mater. Sci. Mater. Electron. 23 (2012) 2152-2156.

[62] S. Santra and A. Paul, Philos. Mag. Lett. 92 (2012) 373-383.

[63] S. Santra, A. Mondal and A. Paul, Metall. Mater. Trans. 43 (2012) 791-795.

[64] V.D. Divya, U. Ramamurty and A. Paul, J. Mater. Res. 26 (2011) 2384-2393.

[65] V.D. Divya, U. Ramamurty and A. Paul, Defect, Diffus. Forum 312-315 (2011) 466-471.

[66] R. Ravi and A. Paul, Intermetallics 19 (2011) 426-428.

[67] V.D. Divya, S.S.K. Balam, U. Ramamurty and A. Paul, Scripta Mater. 62 (2010) 621-624.

[68] V.D. Divya, U. Ramamurty and A. Paul, Intermetallics 18 (2010) 259-266.

[69] P Kiruthika, SK Makineni, C Srivastava, K Chattopadhyay, A Paul, Acta Materialia 105 (2017) 512.

[70] C.R. Kao and Y.A. Chang, Intermetallics 1 (1993) 237-250.

[71] St. Frank, S.V. Divinski, U. Södervall and Chr. Herzig, Acta Mater. 49 (2001) 1399-1411.

[72] C. Cserhati, U. Ugaste, M.J.H. van Dal, N. Lousberg, A.A. Kodentsov and F.J.J. van Loo, Defect Diffus. Forum 194-199 (2001) 189-194.

[73] D. Divya, U. Ramamurty and A. Paul, Philos. Mag. 92 (2012) 2187-2214.

[74] C. M. F. Rae and R. C. Reed, Acta mater. 49 (2001) 4113–4125.

[75] W.J. Cheng, C. J. Wang, Surf.  Coat. Technol. 204 (2009) 824–828.

[76] S. Kobayashi, T. Yakou, Mater. Sci. Eng. A 338 (2002) 44-53.

[77] Y.Y. Chang, C.C. Tsaur, J.C. Rock, Surf. Coat. Technol. 200 (2006) 6588-6593.

[78] C.J. Wang, S.M. Chen, Surf. Coat. Technol. 200 (2006) 6601-6606.

[79] H. Glasbrenner, H.U. Borgstedt, J. Nucl. Mater. 212–215 (1994) 1561-1565.

[80] A.Hayashi, C.R. Kao, Y.A. Chang, Sctipta Mater. 37 (1997) 393-398.

[81] J.H. Gülpen, A.A. Kodentsov, F.J.J. van Loo, Z. Metallkd. 86 (1995) 530-539.

[82] K. Bhanumurthy, W. Krauss, J. Konys, Fusion Sci. Technol., 65（2014) 262-272.

[83] B . A . P i n t , Y . Z h a n g , P.F.Tortorelli, J.A.Haynes and I.G.Wright, Mater. High Temp. 18 (2001) 185-192.